\documentclass[reprint,aps,superscriptaddress,amsmath,amssymb,longbibliography]{revtex4-1}
\usepackage{graphicx}
\usepackage{dcolumn}
\usepackage{bm}
\usepackage{color}
\usepackage{xspace}            
\usepackage{ulem}
\usepackage[colorlinks=true,urlcolor=blue,citecolor=blue,linkcolor=blue,bookmarks=false,
pdfstartview={FitH}]{hyperref}
\usepackage{gensymb}

\begin{document}

\title{Magneto-elastic coupling in Fe-based superconductors}     
\author{S.-F. Wu}\email{sfwu@iphy.ac.cn}
\affiliation{Department of Physics and Astronomy, Rutgers University,
Piscataway, NJ 08854, USA}
\affiliation{Beijing National Laboratory for Condensed Matter
Physics, and Institute of Physics, Chinese Academy of Sciences,
Beijing 100190, China}
\affiliation{School of Physical Sciences, University of Chinese
Academy of Sciences, Beijing 100190, China}
\author{W.-L. Zhang}         
\affiliation{Department of Physics and Astronomy, Rutgers University,
Piscataway, NJ 08854, USA}
\author{V. K. Thorsm{\o}lle}        
\affiliation{Department of Physics and Astronomy, Rutgers University,
Piscataway, NJ 08854, USA}
\author{G. F. Chen}    
\affiliation{Beijing National Laboratory for Condensed Matter
Physics, and Institute of Physics, Chinese Academy of Sciences,
Beijing 100190, China}
\affiliation{Collaborative Innovation Center of Quantum Matter,
Beijing, China}
\author{G. T. Tan}    
\affiliation{Center for Advanced Quantum Studies and Department of
Physics, Beijing Normal University, Beijing 100875, China}
\author{P. C. Dai}    
\affiliation{Center for Advanced Quantum Studies and Department of
Physics, Beijing Normal University, Beijing 100875, China}
\affiliation{Department of Physics and Astronomy, Rice University,
Houston, Texas 77005, USA}
\author{Y. G. Shi}     
\affiliation{Beijing National Laboratory for Condensed Matter
Physics, and Institute of Physics, Chinese Academy of Sciences,
Beijing 100190, China}
\author{C. Q. Jin}    
\affiliation{Beijing National Laboratory for Condensed Matter
Physics, and Institute of Physics, Chinese Academy of Sciences,
Beijing 100190, China}
\affiliation{Collaborative Innovation Center of Quantum Matter,
Beijing, China}
\author{T. Shibauchi}                
\affiliation{Department of Advanced Materials Science, University of
Tokyo, Kashiwa, Chiba 277-8561, Japan}
\author{S. Kasahara}     
\affiliation{Department of Physics, Kyoto University, Sakyo-ku, Kyoto
606-8502, Japan}
\author{Y. Matsuda}
\affiliation{Department of Physics, Kyoto University, Sakyo-ku, Kyoto
606-8502, Japan}
\author{A. S. Sefat}     
\affiliation{Materials Science \& Technology Division, Oak Ridge
National Laboratory, Oak Ridge, TN 37831}
\author{H. Ding}
\affiliation{Beijing National Laboratory for Condensed Matter
Physics, and Institute of Physics, Chinese Academy of Sciences,
Beijing 100190, China}
\affiliation{School of Physical Sciences, University of Chinese
Academy of Sciences, Beijing 100190, China}
\affiliation{Collaborative Innovation Center of Quantum Matter,
Beijing, China}
\author{P. Richard}\email{pierre.richard.qc@gmail.com}
\affiliation{Beijing National Laboratory for Condensed Matter
Physics, and Institute of Physics, Chinese Academy of Sciences,
Beijing 100190, China}
\affiliation{School of Physical Sciences, University of Chinese
Academy of Sciences, Beijing 100190, China}
\affiliation{Collaborative Innovation Center of Quantum Matter,
Beijing, China}
\author{G. Blumberg}\email{girsh@physics.rutgers.edu}
\affiliation{Department of Physics and Astronomy, Rutgers University,
Piscataway, NJ 08854, USA}
\affiliation{National Institute of Chemical Physics and Biophysics,
12618 Tallinn, Estonia}
\date{\today}

\begin{abstract}                                             
We used polarization-resolved Raman scattering to study the
magneto-elastic coupling in the parent compounds of several families
of Fe-based superconductors (BaFe$_2$As$_2$, EuFe$_2$As$_2$, NaFeAs,
LiFeAs, FeSe and LaFeAsO). We observe an emergent $A_g$-symmetry As
phonon mode in the $XY$ scattering geometry whose intensity is
significantly enhanced below the magneto-structural transition only
for compounds showing magnetic ordering. We conclude that the small
lattice anisotropy is insufficient to induce the in-plane electronic
polarizability anisotropy necessary for the observed phonon intensity
enhancement, and interpret this enhancement below the N\'{e}el
temperature in terms of the anisotropy of the magnetic moment and
magneto-elastic coupling. We evidence a Fano line-shape in the $XY$
scattering geometry resulting from a strong coupling between the
$A_g$(As) phonon mode and the $B_{2g}$ symmetry-like electronic
continuum. Strong electron-phonon coupling may be relevant to
superconductivity.

\end{abstract}
                           
               
\maketitle

The lattice, orbital and magnetic degrees of freedom are strongly
coupled in the Fe-based superconductors. This is best evidenced by
the observation, in most parent compounds, of a magnetic transition
from paramagnetic to collinear antiferromagnetic (AFM), occurring at
a temperature $T_N$ slightly lower than the temperature $T_S$ at
which a structural transition from tetragonal to orthorhombic phase
occurs upon cooling. The interplay between these degrees of freedom
is complex and led to a chicken-egg problem for which there is still
no consensual view \cite{Fernandes2014NatPhy,Fernandes2010PRL}. The
electronic structure is directly affected by the structural and
magnetic transitions, notably through nematic transport
properties~\cite{Chu824science2010,Chu710science2012,Kuo958science2016},
as well as by an electronic band folding accompanied by the formation
of a spin-density-wave
gap~\cite{Hu2008PRL,Ran_PRB79,RichardPRL2010,Zhang2016PRB}.

The As height and the related $\textrm{Fe-As-Fe}$ angle are widely
believed to play crucial roles in shaping the magnetic and electronic
properties of the Fe-based superconductors
\cite{Kuroki2009PRB,Lee2014NJP,Baledent2015PRL,Vildosola2008PRB,Bascones2009PRB,YinPRL2008,Yndurain2011EPL,Cruz2010PRL,Zhang2014PRL,Lee2008JPSJ,Zhao2008,KurokiPRB2009,Garbarino2011EPL,Bascones2013PRB}.
Both parameters are modulated by the $c$-axis motion of the As
atom corresponding to a fully symmetric phonon mode ($A_{1g}$)
\cite{Mansart2009PRB,Kim2012NatMat,Avigo2013JPCM,Yang2014PRL,Gerber2015NatCom,Rettig2015PhysRevLett114,Mandal2014PRB}.
First-principles calculations show that the inclusion of the Fe spin
ordering in the calculation of the As phonon mode frequency allows a
good agreement with the energy of the As phonon density-of-states
measured by neutron scattering~
\cite{Yildirim2009PhysicaC,BoeriPRB2010,Zbiri2009PRB,ReznikPRB2009,Hahn2009PRB,Mittal2013PRB,Hahn2013PRB},
suggesting significant magneto-elastic coupling.

Raman scattering can directly probe the As phonon behavior upon
cooling across the magneto-structural transitions. 
As a signature of
the magneto-elastic coupling, a finite intensity of the As phonon in
nearly forbidden scattering geometries below the
magneto-structural transition has been reported in CaFe$_2$As$_2$
\cite{Choi2008PhysRevB78}, EuFe$_2$As$_2$~\cite{ZhangWL2014arxiv},
Ba(Fe$_{1-x}$Co$_{x}$)$_2$As$_2$~\cite{Chauvilere_PRB80,Kretzschmar2016NatPhy,Sugai2012JPSJ},
Ba(Fe$_{1-x}$Au$_{x}$)$_2$As$_2$~\cite{Au_paper},
LaFeAsO~\cite{Kaneko2017arxiv}. 
In particular, the phonon shows
an asymmetric line-shape below $T_N$ in
Ba(Fe$_{1-x}$Co$_{x}$)$_2$As$_2$, suggesting strong magneto-elastic
coupling~\cite{Chauvilere_PRB80,Kretzschmar2016NatPhy}. However, the
details behind this behavior, and its possible link to
superconductivity, have not been stated
satisfactorily. It has long been suggested that the electron-electron
correlations enhance the electron-phonon coupling in the Fe-based
superconductors and that the fully symmetric As vibration is related
to the superconducting
properties~\cite{Mandal2014PRB,Garbarino2011EPL,Egami2010JAdvances,Gerber2017Science}.

In this Letter, we use polarized Raman scattering to study the
temperature dependence of the magneto-elastic coupling for the fully
symmetric phonon associated with the $c$-axis motion of the As atom
for typical ``122", ``111", ``1111" and ``11" systems of Fe-based
superconductors. 
For all compounds showing magnetic ordering, we
observe strong intensity for the fully-symmetric As mode appearing
below $T_N$ in the nearly forbidden $XY$ scattering channel as a
result of significantly enhanced anisotropy of the in-plane
electronic polarizability, while no such enhancement is found for 
compounds without magnetically-ordered state. 
Because the lattice anisotropy
$\delta=(a-b)/(a+b)$ below $T_S$ is relatively small, we conclude
that magneto-elastic coupling below $T_N$ is essential. We interpret
the $A_g$(As) phonon intensity enhancement below $T_N$ in terms of
strong coupling to the anisotropic in-plane magnetic moment. The
study of the polarization dependence of the As phonon suggests that
the mode is coupled to the non-symmetric $XY$-like electronic
continuum. The asymmetric line-shape of $A_g$(As) phonon is described
by a Fano model with a magneto-elastic coupling constant proportional
to the magnetic order parameter. As the coupling between the
$XY$-like electronic continuum and magnetism may survive in the
superconducting compounds, our results emphasize the role played by
the electron-phonon coupling in enhancing $T_c$.

Single crystals of materials listed in Table~\ref{TSTN} were grown as
described in
Refs.~\cite{Sefat2013bulk,Li2015PRB,Tanatar2012PRB,Kamihara2008JACS,Clarina2008Nature,Hosoi2016PNAS}.
The corresponding structural phase transition temperature ($T_S$) and
magnetic phase transition temperature ($T_N$) are summarized in
Table~\ref{TSTN}.
Raman measurements on BaFe$_{2}$As$_2$, NaFeAs, EuFe$_2$As$_2$,
LiFeAs, FeSe were performed using the spectrometer described in
Refs.~\cite{ZhangWL2014arxiv,Thorsmolle2016PRB}. The measurements on
LaFeAsO were performed in a back scattering geometry using a T64000
triple-stage spectrometer.

\begin{table}[!t]
\caption{\label{TSTN} Summary of $T_{S}$, $T_{N}$(in Kelvin), lattice
orthorhombicity~($\delta=(a-b)/(a+b)$), intensity ratio of $Ag$
phonon in $XY$ and $XX$ geometries, and ordered magnetic moment/Fe M
(in $\mu_B$) for compounds studied in this manuscript.}
\begin{ruledtabular}
\begin{tabular}{ccccc}
Sample&$T_{S}$/$T_{N}$&$\delta$ (\%)&$I_{XY}/I_{XX}$&M\\
\hline  
EuFe$_2$As$_2$~\cite{Sefat2013bulk}&175/175&0.5~\cite{Marcus2008JPCM}&3.3&0.98
\cite{Xiao2009EuFe2As2}\\
BaFe$_2$As$_2$~\cite{Sefat2013bulk}&135/135&0.4~\cite{Huang2008PRL}&3.1&0.87
\cite{Dai2015RevModPhys}\\
NaFeAs~\cite{Tanatar2012PRB}&55/40&0.18~\cite{Li2009PRB}&0.16&0.09
\cite{Dai2015RevModPhys}\\
LaFeAsO~\cite{Kamihara2008JACS,Clarina2008Nature}&155/137&0.24~\cite{delaCruz2008}&0.54&0.36-0.6
\cite{Dai2015RevModPhys}\\
\hline
FeSe~\cite{Hosoi2016PNAS}&90/-&0.25~\cite{McQueen2009PRL}&0.017&-\\
\hline
LiFeAs~\cite{WangXC2010LiFeAs}&-/-&0&0&-\\   
\end{tabular}
\end{ruledtabular}
\begin{raggedright}
\end{raggedright}
\end{table}

The phononic Raman scattering intensity is proportional to
$I\propto|\hat{e}_i\cdot \textbf{R}\cdot\hat{e}_s|^2$, where
$\hat{e}_i$ and $\hat{e}_s$ are the polarization unit vectors of the
incoming and scattering light, respectively, and $\textbf{R}$ is the
Raman tensor~\cite{CardonaBookI}. For the $D_{4h}$ point group the
$XX$, $XY$, $X'X'$ and $X'Y'$ polarization geometries probe $A_{1g} +
B_{1g}$, $A_{2g}+ B_{2g}$, $A_{1g} + B_{2g}$ and $A_{2g} + B_{1g}$
symmetry excitations, respectively. In the orthorhombic phase with
$D_{2h}$ point group symmetry, the unit cell rotates by 45\degree;
the $A_{1g}$ and $B_{2g}$ representations of the $D_{4h}$ point group
merge into the $A_g$ representation of the $D_{2h}$ point group, and
$A_{2g}$ and $B_{1g}$ ($D_{4h}$) merge into $B_{1g}$ ($D_{2h}$). In
the orthorhombic phase, the $XX$ and $XY$ polarization geometries
probe $A_g + B_{1g}$ and $A_g$ symmetry excitations,
respectively~\cite{Zhang2016PRB}.

\begin{figure}[t]   
\begin{center}
\includegraphics[width=\columnwidth]{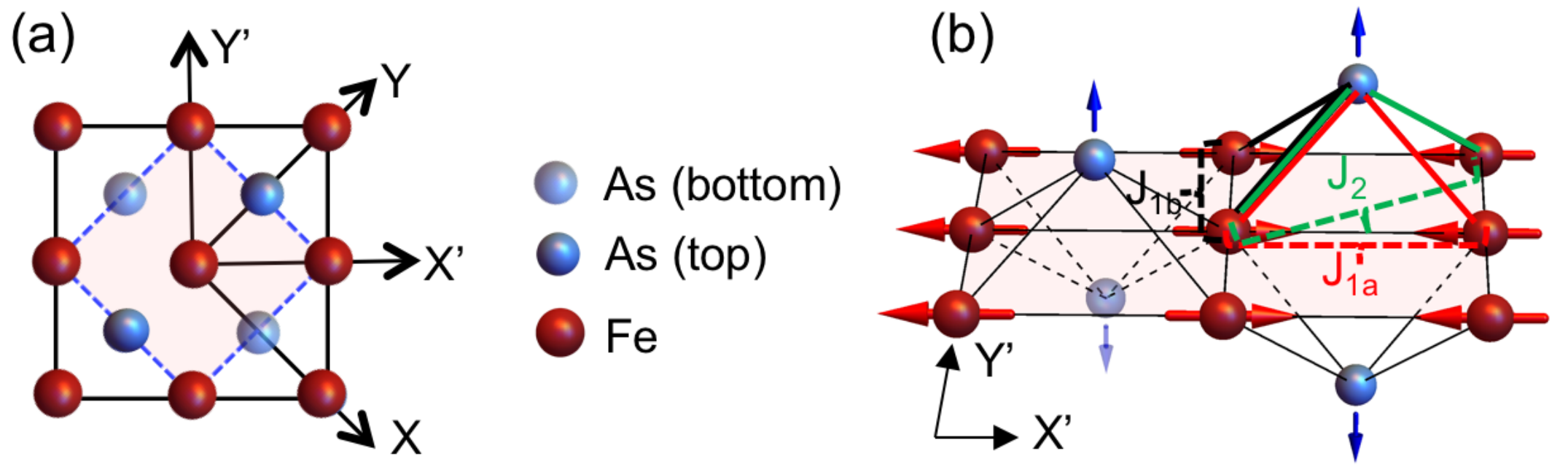}
\end{center}
\caption{\label{Fig1_structure}(Color online) (a) Definition of the
crystallographic directions in the tetragonal 2-Fe unit cell above
$T_S$ (light red shaded area) and 4-Fe orthorhombic magnetic unit
cell below $T_N$ (black solid lines). (b) Schematic diagram of the
magnetic structure. Red arrows: Fe local moments forming collinear
AFM order. Blue arrows: $c$-axes vibrations of the fully symmetric As
phonon mode. The red and black solid lines illustrate the
super-exchange paths of the nearest Fe neighbors, $J_{1a}$ and
$J_{1b}$. The green solid lines illustrate the super-exchange path of
the next-nearest Fe neighbors, $J_{2}$.}    
\end{figure}

Before investigating the behavior of the $A_{1g}/A_g$ symmetry As
phonon across the magneto-structural transitions, we first examine 
the $A_{1g}$ and $A_g$ Raman tensors: 
\begin{displaymath}
\textrm{ $A_{1g}^{D_{4h}}$=}
\left(\begin{array}{ccc}
\bar{a} & 0 &0\\
0 & \bar{a} &0\\
0 & 0 &\bar{c}
\end{array}\right),
\textrm{ $A_g^{D_{2h}}$=}
\left(\begin{array}{ccc}
\frac{(\bar{a}'+\bar{b}')}{2} & \frac{(\bar{a}'-\bar{b}')}{2}  &0\\
\frac{(\bar{a}'-\bar{b}')}{2} &\frac{(\bar{a}'+\bar{b}')}{2} &0\\
0 & 0 &\bar{c}
\end{array}\right)
\end{displaymath}  
\noindent where $A_g^{D_{2h}}$ (orthorhombic phase) has been rotated
by 45\degree~to keep the same $XYZ$ axis notation as in the
tetragonal phase. $\bar{a}'$ and $\bar{b}'$ are the diagonal elements
of the $A_g^{D_{2h}}$ Raman tensor in the natural coordinate system
of the orthorhombic phase (before the 45\degree~rotation).  
                                                    
\begin{figure*}[t]   
\begin{center}
\includegraphics[width=2\columnwidth]{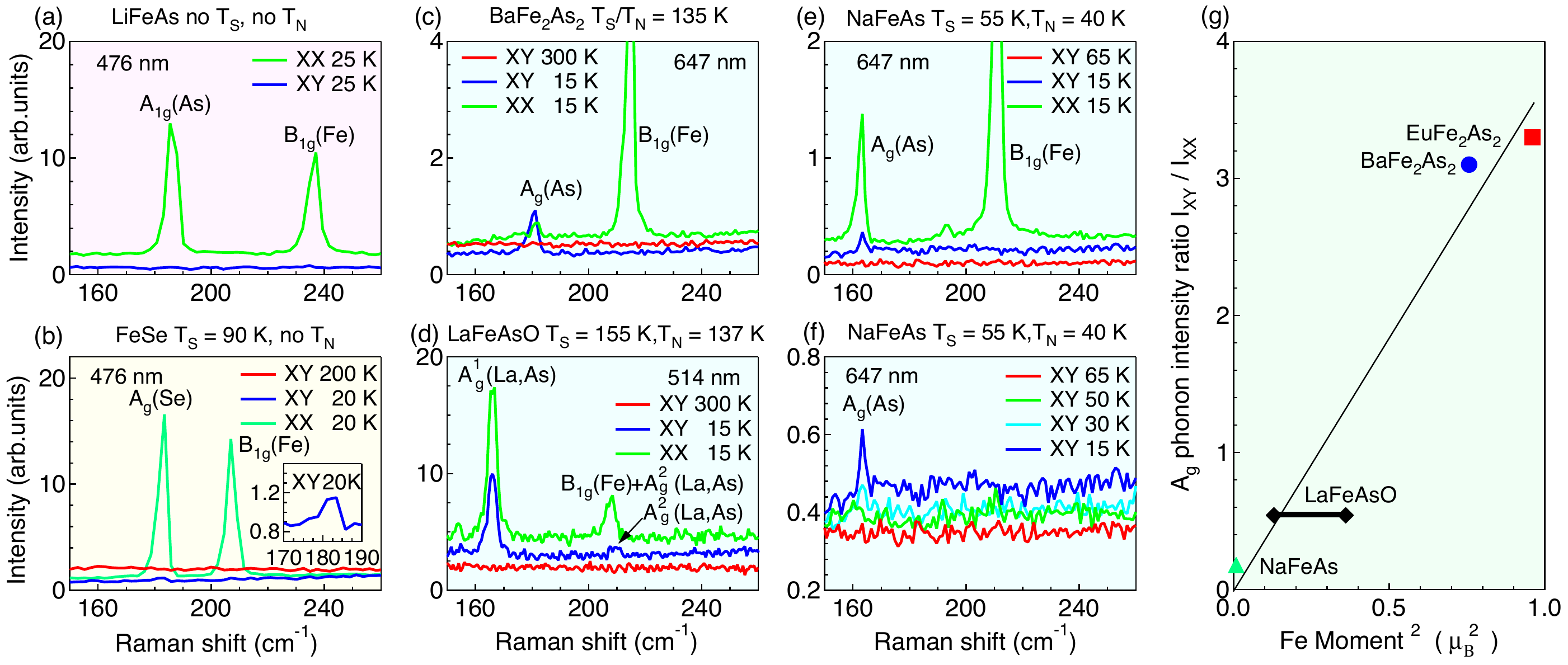}
\end{center}
\caption{\label{Fig2_Ag_leakage} 
Comparison of Raman
spectra (shifted for clarity) for the $XX$ and $XY$ scattering
geometries for different parent compounds: 
(a) LiFeAs, (b) FeSe, (c)
BaFe$_{2}$As$_2$, (d) LaFeAsO, (e) and (f) NaFeAs. 
When finite, the
$T_N$ and $T_S$ values are indicated on the top of the corresponding
panel. (g) $A_g$(As) phonon intensity ratio in the $XY$ and $XX$
geometries ($I_{XY}/I_{XX}$) as a function of the squared ordered
magnetic moment/Fe~\cite{Xiao2009EuFe2As2,Dai2015RevModPhys}. 
The black line is a linear fit.} 
\end{figure*}

Accordingly, the $A_{1g}$-symmetry mode is forbidden in the $XY$
scattering geometry in the tetragonal phase. This is the case for
LiFeAs, which shows no structural nor magnetic transition. As shown
in Fig. \ref{Fig2_Ag_leakage}(a), sharp Raman phonon peaks at
186~cm$^{-1}$ and 237~cm$^{-1}$, corresponding to a $A_{1g}$(As) and
a $B_{1g}$(Fe) modes, respectively, are detected in the $XX$
scattering geometry. However, as expected for the tetragonal
structure of LiFeAs, these modes have no intensity in the $XY$
scattering geometry.

If anisotropy develops in the orthorhombic phase, the $A_{g}$ anion
mode may acquire a finite intensity $|(\bar{a}'-\bar{b}')/2|$$^2$ in
the $XY$ scattering geometry related to the anisotropy of the
in-plane polarizability associated to this $A_{g}$ anion mode because
$\bar{a}'$ and $\bar{b}'$ are the polarizability derivatives along
the two Fe-Fe orthogonal directions ($X'$ and $Y'$) in the
orthorhombic phase. Since the lattice orthorhombicity $\delta$ is
small (Table~\ref{TSTN}), the intensity is expected to be weak. For
example, for the FeSe material, which exhibits a structural
transition at 90~K \cite{BohmerPRB87,BohmerPRL114} but no long-range
magnetic ordering, we observe a $A_g$(Se) phonon at 180~cm$^{-1}$ and
a $B_{1g}$(Fe) phonon at 208 cm$^{-1}$ for the $XX$ polarization.
Although the intensity of the $A_g$(Se) phonon with the $XY$
polarization is finite at 20~K, it is only 2\% of the corresponding
intensity recorded for the $XX$ polarization~[Table \ref{TSTN}].

In contrast, BaFe$_2$As$_2$ with magnetic ordering clearly shows the
181~cm$^{-1}$ $A_{g}$ (As)
mode~\cite{Litvinchuk2008PhysRevB,Chauviere2009PhysRevB80,Sugai2012JPSJ,Rahlenbeck2009PRB}
in the $XY$ scattering geometry below $T_N$
[Fig.~\ref{Fig2_Ag_leakage}(c)]. 
Similar observation is made for
NaFeAs [Figs.~\ref{Fig2_Ag_leakage}(e) and~\ref{Fig2_Ag_leakage}(f)],
which also encounters both a structural and a magnetic transition:
(i) We observe only a weak intensity between $T_S$ and $T_N$, and 
(ii) the 162~cm$^{-1}$ $A_g$(As) phonon mode appears in the $XY$
spectra
only below $T_N$.
LaFeAsO~\cite{Kaneko2017arxiv,Hadjiev2008PhysRevB,Zhao2009SCT}
(Fig.~\ref{Fig2_Ag_leakage}(d))
is another system with split $T_S$ and $T_N$ transitions. In this
case as well, we detect sizable intensity for the $A^1_g$(in-phase La
and As) mode at 166~cm$^{-1}$ and the $A^2_g$(out-of-phase La and As)
mode at 209~cm$^{-1}$  in the $XY$ scattering geometry below $T_N$
[Fig.~\ref{Fig2_Ag_leakage}(d)].     

To quantify the intensity of the $A_g$(As) phonon in the $XY$
scattering geometry below $T_N$ in different families of Fe-based
superconductors, we study the ratio between the $A_g$(As) peak
intensity in the $XY$ and $XX$ scattering geometries $I_{XY}/I_{XX}$.
This ratio is proportional to
$|(\bar{a}'-\bar{b}')/(\bar{a}'+\bar{b}')|$$^2$, which is a direct
measure of the in-plane polarizability anisotropy of the $A_{g}$(As)
mode.
Based on Table~\ref{TSTN}, the ratio $I_{XY}/I_{XX}$ is significant
only for compounds with long-range magnetic ordering. For example,
the ratio $I_{XY}/I_{XX}$ is 300\% for BaFe$_2$As$_2$, 16\% for
NaFeAs and 50\% for LaFeAsO, as compared to 2\% for FeSe, \textit{i.
e.} 1 to 2 orders of magnitude smaller. Such behavior cannot be
solely explained by weak lattice orthorhombicity $\delta$, and
indicates that the intensity of the $A_g$(As) phonon in the $XY$
scattering geometry is mainly controlled by the magneto-elastic
coupling, for which we argue that the strength originates from the
anisotropy of the magnetic interactions in the Fe-As plane that are
modulated by the $c$-axis motion of the As atoms. We can estimate the
strength of the magneto-elastic coupling by comparing the
$I_{XY}/I_{XX}$ intensity ratios in the magnetically-ordered
compounds to that in FeSe. As compared to FeSe, the coupling strength
values are 200, 25 and 8 in BaFe$_2$As$_2$, LaFeAsO and NaFeAs,
respectively. 

In Fig.~\ref{Fig2_Ag_leakage}(g), we show that the $I_{XY}/I_{XX}$
ratio of the $A_g$(As) phonon intensity for different Fe-based
materials scales linearly with the square of the magnetic moment $M$,
indicating that the magneto-elastic coupling constant is proportional
to the ordered magnetic moment $M$. In Ref.~\cite{Bascones2013PRB},
the magneto-elastic coupling was explicitly calculated within a
tight-binding Slater-Koster formalism. The study predicts a large
enhancement of the As mode in the $XY$ scattering geometry,
consistent with experimental observation
[Fig.~\ref{Fig2_Ag_leakage}], due to the anisotropy of the
Slater-Koster energy integrals in the magnetically ordered state.

\begin{figure}[t] 
\begin{center}
\includegraphics[width=\columnwidth]{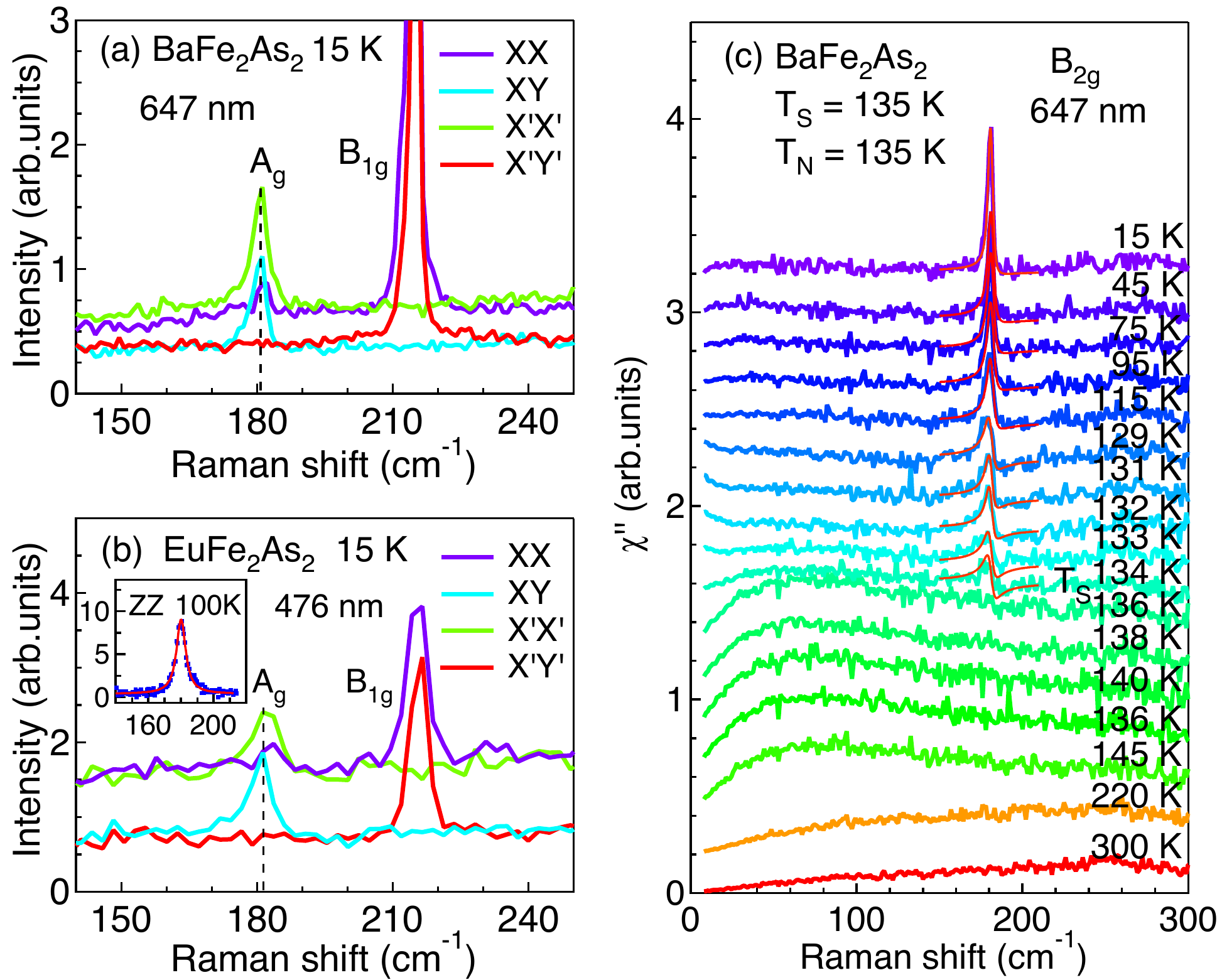}
\end{center}
\caption{\label{Fig3_BaFe2As2_4pol_EuFe2As2_15K}(Color online) Raman
spectra in the $XX$, $XY$, $X'X'$ and $X'Y'$ geometries for (a)
BaFe$_2$As$_2$ and (b) EuFe$_2$As$_2$ at 15 K. 
The inset in (b) shows spectra for $ZZ$ 
polarization from a polished $ac$ surface
of EuFe$_2$As$_2$ for the in the magnetic
state at 100~K excited with 752 nm laser
line~\cite{Zhang2016PRB94}. 
The red line is a fit of the $A_g$ phonon with a Lorentzian function. 
(c) $T$-dependence of Raman spectra for BaFe$_2$As$_2$
in the $XY$ scattering geometry (shifted for clarity). 
The solid red lines are Fano-shape fits~\cite{Au_paper}. The
spectral resolution is about 0.85\,cm$^{-1}$.}
\end{figure}

Above we have established that the intensity enhancement of the
$A_g$(As) phonon mode in the $XY$ scattering geometry depends on the
presence of ordered magnetic moment. 
We now address the coupling between the $A_g$(As) phonon and the
$B_{2g}$-like electronic continuum below $T_N$. 
In Figs.~\ref{Fig3_BaFe2As2_4pol_EuFe2As2_15K}(a-b) we present the
polarization
dependence of the spectra for BaFe$_2$As$_2$ and EuFe$_2$As$_2$ at
15~K.
The line-shape of the  $A_g$(As) phonon in the $XX$ and $ZZ$ 
scattering geometries is symmetric, in contrast to asymmetric 
interference Fano shape observed in the $XY$ and $X'X'$ 
geometries~\cite{Fano1961PhysRev,CardonaBookI}.  
The polarization analysis suggests that interfering with the phonon 
electronic continuum must have $B_{2g}$ symmetry.  
A  $B_{2g}$-like continuum is allowed to couple to $A_{g}$(As) phonon 
in the orthorhombic phase because below the $D_{4h}\rightarrow 
D_{2h}$ transition the 
$A_{1g}$ and $B_{2g}$ representations merge into the $A_g$
irreducible representation. 
For $XY$ and $X'X'$ geometry, where the
$B_{2g}$-like continuum is represent, the bare $A_g$(As) phonon is
coupled to the $B_{2g}$-like electronic continuum, giving rise to an
asymmetric Fano line-shape.

In addition, we argue that
the density-of-states of the $B_{2g}$-like continuum is 
temperature-dependent. 
In Fig.~\ref{Fig3_BaFe2As2_4pol_EuFe2As2_15K}(c), we show 
temperature evolution of the Raman spectra in $XY$ scattering 
geometry for BaFe$_{2}$As$_2$. 
Just below $T_N$ the $A_g$(As) phonon instantly appears 
with a visibly asymmetric line-shape.
The mode sharpens and becomes more symmetric upon cooling,
which we attribute to a decrease in the electronic density-of-states
at the Fermi level, contributing to the $B_{2g}$-like continuum, due 
to the spin-density-wave gap 
formation~\cite{Hu2008PRL,Ran_PRB79,RichardPRL2010,Zhang2016PRB}.

To quantify the electron-phonon interaction, we constructed a Fano 
model (Eq.~3 in Ref.~\cite{Au_paper}) with a magneto-elastic coupling
constant 
proportional to the magnetic ordered moment. 
The Fano interference is between the $A_g$(As) phonon mode and
the $B_{2g}$-like electronic continuum below $T_N$. 
Such model well describes the data for BaFe$_2$As$_2$, 
Fig.~\ref{Fig3_BaFe2As2_4pol_EuFe2As2_15K}(c).  
Same approach also describes all spectral features for 
Ba(Fe$_{1-x}$Au$_x$)$_2$As$_2$~\cite{Au_paper}. 
Hence, the polarization-resolved Raman spectroscopy implemented here
represents an all-inclusive tool to study the magnetism and the
magneto-elastic interaction in the Fe-based superconductors.

Finally, we discuss the implications of the electron-phonon coupling
to superconductivity. 
Early calculations show that when magnetic moments are
included~\cite{Boeri2010PRB82}, the electron-phonon coupling is
enhanced by 50\% as compared with non-magnetic
calculations~\cite{Boeri2008PRL101}. 
According to more recent calculations, the electron-phonon matrix
element is rather enhanced four times in the ordered AFM state due to
the presence of a $d_{xz}/d_{yz}$ Fermi surface near the zone
corner~\cite{Coh2016PRB,Coh2015NJP}. In particular, the Eliashberg
spectral function $\alpha^2F$ is enhanced 4 times around 22~meV,
which corresponds to the $A_{1g}$ mode
energy~\cite{Coh2016PRB,Coh2015NJP}.  
Thus, one cannot rule out the possibility that the enhanced intraband
electron-phonon coupling in the AFM phase, if sufficiently large,
could enhance the paring temperature.

In conclusion, we revealed a significant intensity enhancement of the
emergent $A_g$(As) phonon mode in the $XY$ scattering geometry below
$T_N$ only for parent compounds of Fe-based superconductors showing
magnetic order. We argue that the in-plane electronic polarizability
anisotropy necessary for the $A_g$(As) phonon intensity enhancement
originates from the anisotropy of the magnetic interactions in the
Fe-As plane that are modulated by the $c$-axis motion of the As
atoms. In particular, we demonstrate a magneto-elastic coupling
between the $A_g$(As) phonon and the $B_{2g}$-like electronic
continuum that is essential to $A_g$(As) phonon intensity
enhancement. The asymmetric line-shape of $A_g$(As) phonon is well
explained by a Fano model with a magneto-elastic coupling constant
proportional to the ordered magnetic moment. Our results identify
strong electron-phonon coupling in the magnetic phase of Fe-based
superconductors, which could enhance the paring temperature.

We thank E. Bascones and K. Haule for discussions.
The research at Rutgers was supported by the US Department of Energy,
Basic Energy
Sciences, and Division of Materials Sciences and Engineering under
Grant No. DE-SC0005463. 
The work at ORNL was supported by the US
Department of Energy, Basic Energy Sciences, Materials Sciences and
Engineering Division. Work at IOP was supported  by grants  from NSFC
(11674371,11274362,11774399 and 11474330) and  MOST (2015CB921301,
2016YFA0401000 and 2016YFA0300300) of China.


\begin{thebibliography}{76}%
\makeatletter
\providecommand \@ifxundefined [1]{%
 \@ifx{#1\undefined}
}%
\providecommand \@ifnum [1]{%
 \ifnum #1\expandafter \@firstoftwo
 \else \expandafter \@secondoftwo
 \fi
}%
\providecommand \@ifx [1]{%
 \ifx #1\expandafter \@firstoftwo
 \else \expandafter \@secondoftwo
 \fi
}%
\providecommand \natexlab [1]{#1}%
\providecommand \enquote  [1]{``#1''}%
\providecommand \bibnamefont  [1]{#1}%
\providecommand \bibfnamefont [1]{#1}%
\providecommand \citenamefont [1]{#1}%
\providecommand \href@noop [0]{\@secondoftwo}%
\providecommand \href [0]{\begingroup \@sanitize@url \@href}%
\providecommand \@href[1]{\@@startlink{#1}\@@href}%
\providecommand \@@href[1]{\endgroup#1\@@endlink}%
\providecommand \@sanitize@url [0]{\catcode `\\12\catcode `\$12\catcode
  `\&12\catcode `\#12\catcode `\^12\catcode `\_12\catcode `\%12\relax}%
\providecommand \@@startlink[1]{}%
\providecommand \@@endlink[0]{}%
\providecommand \url  [0]{\begingroup\@sanitize@url \@url }%
\providecommand \@url [1]{\endgroup\@href {#1}{\urlprefix }}%
\providecommand \urlprefix  [0]{URL }%
\providecommand \Eprint [0]{\href }%
\providecommand \doibase [0]{http://dx.doi.org/}%
\providecommand \selectlanguage [0]{\@gobble}%
\providecommand \bibinfo  [0]{\@secondoftwo}%
\providecommand \bibfield  [0]{\@secondoftwo}%
\providecommand \translation [1]{[#1]}%
\providecommand \BibitemOpen [0]{}%
\providecommand \bibitemStop [0]{}%
\providecommand \bibitemNoStop [0]{.\EOS\space}%
\providecommand \EOS [0]{\spacefactor3000\relax}%
\providecommand \BibitemShut  [1]{\csname bibitem#1\endcsname}%
\let\auto@bib@innerbib\@empty
\bibitem [{\citenamefont {Fernandes}\ \emph {et~al.}(2014)\citenamefont
  {Fernandes}, \citenamefont {Chubukov},\ and\ \citenamefont
  {Schmalian}}]{Fernandes2014NatPhy}%
  \BibitemOpen
  \bibfield  {author} {\bibinfo {author} {\bibfnamefont {R.~M.}\ \bibnamefont
  {Fernandes}}, \bibinfo {author} {\bibfnamefont {A.~V.}\ \bibnamefont
  {Chubukov}}, \ and\ \bibinfo {author} {\bibfnamefont {J.}~\bibnamefont
  {Schmalian}},\ }\bibfield  {title} {\enquote {\bibinfo {title} {What drives
  nematic order in iron-based superconductors?}}\ }\href
  {http://dx.doi.org/10.1038/nphys2877} {\bibfield  {journal} {\bibinfo
  {journal} {Nat. Phys.}\ }\textbf {\bibinfo {volume} {10}},\ \bibinfo {pages}
  {97} (\bibinfo {year} {2014})}\BibitemShut {NoStop}%
\bibitem [{\citenamefont {Fernandes}\ \emph {et~al.}(2010)\citenamefont
  {Fernandes}, \citenamefont {VanBebber}, \citenamefont {Bhattacharya},
  \citenamefont {Chandra}, \citenamefont {Keppens}, \citenamefont {Mandrus},
  \citenamefont {McGuire}, \citenamefont {Sales}, \citenamefont {Sefat},\ and\
  \citenamefont {Schmalian}}]{Fernandes2010PRL}%
  \BibitemOpen
  \bibfield  {author} {\bibinfo {author} {\bibfnamefont {R.~M.}\ \bibnamefont
  {Fernandes}}, \bibinfo {author} {\bibfnamefont {L.~H.}\ \bibnamefont
  {VanBebber}}, \bibinfo {author} {\bibfnamefont {S.}~\bibnamefont
  {Bhattacharya}}, \bibinfo {author} {\bibfnamefont {P.}~\bibnamefont
  {Chandra}}, \bibinfo {author} {\bibfnamefont {V.}~\bibnamefont {Keppens}},
  \bibinfo {author} {\bibfnamefont {D.}~\bibnamefont {Mandrus}}, \bibinfo
  {author} {\bibfnamefont {M.~A.}\ \bibnamefont {McGuire}}, \bibinfo {author}
  {\bibfnamefont {B.~C.}\ \bibnamefont {Sales}}, \bibinfo {author}
  {\bibfnamefont {A.~S.}\ \bibnamefont {Sefat}}, \ and\ \bibinfo {author}
  {\bibfnamefont {J.}~\bibnamefont {Schmalian}},\ }\bibfield  {title} {\enquote
  {\bibinfo {title} {Effects of nematic fluctuations on the elastic properties
  of iron arsenide superconductors},}\ }\href {\doibase
  10.1103/PhysRevLett.105.157003} {\bibfield  {journal} {\bibinfo  {journal}
  {Phys. Rev. Lett.}\ }\textbf {\bibinfo {volume} {105}},\ \bibinfo {pages}
  {157003} (\bibinfo {year} {2010})}\BibitemShut {NoStop}%
\bibitem [{\citenamefont {Chu}\ \emph {et~al.}(2010)\citenamefont {Chu},
  \citenamefont {Analytis}, \citenamefont {De~Greve}, \citenamefont {McMahon},
  \citenamefont {Islam}, \citenamefont {Yamamoto},\ and\ \citenamefont
  {Fisher}}]{Chu824science2010}%
  \BibitemOpen
  \bibfield  {author} {\bibinfo {author} {\bibfnamefont {J.-H.}\ \bibnamefont
  {Chu}}, \bibinfo {author} {\bibfnamefont {J.~G.}\ \bibnamefont {Analytis}},
  \bibinfo {author} {\bibfnamefont {K.}~\bibnamefont {De~Greve}}, \bibinfo
  {author} {\bibfnamefont {P.~L.}\ \bibnamefont {McMahon}}, \bibinfo {author}
  {\bibfnamefont {Z.}~\bibnamefont {Islam}}, \bibinfo {author} {\bibfnamefont
  {Y.}~\bibnamefont {Yamamoto}}, \ and\ \bibinfo {author} {\bibfnamefont
  {I.~R.}\ \bibnamefont {Fisher}},\ }\bibfield  {title} {\enquote {\bibinfo
  {title} {In-plane resistivity anisotropy in an underdoped iron arsenide
  superconductor},}\ }\href {\doibase 10.1126/science.1190482} {\bibfield
  {journal} {\bibinfo  {journal} {Science}\ }\textbf {\bibinfo {volume}
  {329}},\ \bibinfo {pages} {824} (\bibinfo {year} {2010})}\BibitemShut
  {NoStop}%
\bibitem [{\citenamefont {Chu}\ \emph {et~al.}(2012)\citenamefont {Chu},
  \citenamefont {Kuo}, \citenamefont {Analytis},\ and\ \citenamefont
  {Fisher}}]{Chu710science2012}%
  \BibitemOpen
  \bibfield  {author} {\bibinfo {author} {\bibfnamefont {J.~H.}\ \bibnamefont
  {Chu}}, \bibinfo {author} {\bibfnamefont {H.~H.}\ \bibnamefont {Kuo}},
  \bibinfo {author} {\bibfnamefont {J.~G.}\ \bibnamefont {Analytis}}, \ and\
  \bibinfo {author} {\bibfnamefont {I.~R.}\ \bibnamefont {Fisher}},\ }\bibfield
   {title} {\enquote {\bibinfo {title} {Divergent nematic susceptibility in an
  iron arsenide superconductor},}\ }\href {\doibase 10.1126/science.1221713}
  {\bibfield  {journal} {\bibinfo  {journal} {Science}\ }\textbf {\bibinfo
  {volume} {337}},\ \bibinfo {pages} {710} (\bibinfo {year}
  {2012})}\BibitemShut {NoStop}%
\bibitem [{\citenamefont {Kuo}\ \emph {et~al.}(2016)\citenamefont {Kuo},
  \citenamefont {Chu}, \citenamefont {Palmstrom}, \citenamefont {Kivelson},\
  and\ \citenamefont {Fisher}}]{Kuo958science2016}%
  \BibitemOpen
  \bibfield  {author} {\bibinfo {author} {\bibfnamefont {H.~H.}\ \bibnamefont
  {Kuo}}, \bibinfo {author} {\bibfnamefont {J.~H.}\ \bibnamefont {Chu}},
  \bibinfo {author} {\bibfnamefont {J.~C.}\ \bibnamefont {Palmstrom}}, \bibinfo
  {author} {\bibfnamefont {S.~A.}\ \bibnamefont {Kivelson}}, \ and\ \bibinfo
  {author} {\bibfnamefont {I.~R.}\ \bibnamefont {Fisher}},\ }\bibfield  {title}
  {\enquote {\bibinfo {title} {Ubiquitous signatures of nematic quantum
  criticality in optimally doped {F}e-based superconductors},}\ }\href
  {\doibase 10.1126/science.aab0103} {\bibfield  {journal} {\bibinfo  {journal}
  {Science}\ }\textbf {\bibinfo {volume} {352}},\ \bibinfo {pages} {958}
  (\bibinfo {year} {2016})}\BibitemShut {NoStop}%
\bibitem [{\citenamefont {Hu}\ \emph {et~al.}(2008)\citenamefont {Hu},
  \citenamefont {Dong}, \citenamefont {Li}, \citenamefont {Li}, \citenamefont
  {Zheng}, \citenamefont {Chen}, \citenamefont {Luo},\ and\ \citenamefont
  {Wang}}]{Hu2008PRL}%
  \BibitemOpen
  \bibfield  {author} {\bibinfo {author} {\bibfnamefont {W.~Z.}\ \bibnamefont
  {Hu}}, \bibinfo {author} {\bibfnamefont {J.}~\bibnamefont {Dong}}, \bibinfo
  {author} {\bibfnamefont {G.}~\bibnamefont {Li}}, \bibinfo {author}
  {\bibfnamefont {Z.}~\bibnamefont {Li}}, \bibinfo {author} {\bibfnamefont
  {P.}~\bibnamefont {Zheng}}, \bibinfo {author} {\bibfnamefont {G.~F.}\
  \bibnamefont {Chen}}, \bibinfo {author} {\bibfnamefont {J.~L.}\ \bibnamefont
  {Luo}}, \ and\ \bibinfo {author} {\bibfnamefont {N.~L.}\ \bibnamefont
  {Wang}},\ }\bibfield  {title} {\enquote {\bibinfo {title} {Origin of the spin
  density wave instability in {A}{F}e$_2${A}s$_{2}$ ({A}={B}a,{S}r) as revealed
  by optical spectroscopy},}\ }\href {\doibase 10.1103/PhysRevLett.101.257005}
  {\bibfield  {journal} {\bibinfo  {journal} {Phys. Rev. Lett.}\ }\textbf
  {\bibinfo {volume} {101}},\ \bibinfo {pages} {257005} (\bibinfo {year}
  {2008})}\BibitemShut {NoStop}%
\bibitem [{\citenamefont {{Y. Ran, F. Wang, H. Zhai, A. Vishwanath and D.-H.
  Lee}}(2009)}]{Ran_PRB79}%
  \BibitemOpen
  \bibfield  {author} {\bibinfo {author} {\bibnamefont {{Y. Ran, F. Wang, H.
  Zhai, A. Vishwanath and D.-H. Lee}}},\ }\bibfield  {title} {\enquote
  {\bibinfo {title} {{Nodal spin density wave and band topology of the
  FeAs-based materials}},}\ }\href {\doibase 10.1103/PhysRevB.79.014505}
  {\bibfield  {journal} {\bibinfo  {journal} {Phys. Rev. B}\ }\textbf {\bibinfo
  {volume} {79}},\ \bibinfo {pages} {014505} (\bibinfo {year}
  {2009})}\BibitemShut {NoStop}%
\bibitem [{\citenamefont {{P. Richard, K. Nakayama, T. Sato, M. Neupane, Y.-M.
  Xu, J. H. Bowen, G. F. Chen, J. L. Luo, N. L. Wang, X. Dai, Z. Fang, H. Ding
  and T. Takahashi}}(2010)}]{RichardPRL2010}%
  \BibitemOpen
  \bibfield  {author} {\bibinfo {author} {\bibnamefont {{P. Richard, K.
  Nakayama, T. Sato, M. Neupane, Y.-M. Xu, J. H. Bowen, G. F. Chen, J. L. Luo,
  N. L. Wang, X. Dai, Z. Fang, H. Ding and T. Takahashi}}},\ }\bibfield
  {title} {\enquote {\bibinfo {title} {{Observation of {D}irac Cone Electronic
  Dispersion in BaFe$_2$As$_2$}},}\ }\href {\doibase
  10.1103/PhysRevL.104.137001} {\bibfield  {journal} {\bibinfo  {journal}
  {Phys. Rev. Lett.}\ }\textbf {\bibinfo {volume} {104}},\ \bibinfo {pages}
  {137001} (\bibinfo {year} {2010})}\BibitemShut {NoStop}%
\bibitem [{\citenamefont {Zhang}\ \emph
  {et~al.}(2016{\natexlab{a}})\citenamefont {Zhang}, \citenamefont {Yin},
  \citenamefont {Ignatov}, \citenamefont {Bukowski}, \citenamefont {Karpinski},
  \citenamefont {Sefat}, \citenamefont {Ding}, \citenamefont {Richard},\ and\
  \citenamefont {Blumberg}}]{Zhang2016PRB}%
  \BibitemOpen
  \bibfield  {author} {\bibinfo {author} {\bibfnamefont {W.-L.}\ \bibnamefont
  {Zhang}}, \bibinfo {author} {\bibfnamefont {Z.~P.}\ \bibnamefont {Yin}},
  \bibinfo {author} {\bibfnamefont {A.}~\bibnamefont {Ignatov}}, \bibinfo
  {author} {\bibfnamefont {Z.}~\bibnamefont {Bukowski}}, \bibinfo {author}
  {\bibfnamefont {Janusz}\ \bibnamefont {Karpinski}}, \bibinfo {author}
  {\bibfnamefont {Athena~S.}\ \bibnamefont {Sefat}}, \bibinfo {author}
  {\bibfnamefont {H.}~\bibnamefont {Ding}}, \bibinfo {author} {\bibfnamefont
  {P.}~\bibnamefont {Richard}}, \ and\ \bibinfo {author} {\bibfnamefont
  {G.}~\bibnamefont {Blumberg}},\ }\bibfield  {title} {\enquote {\bibinfo
  {title} {Raman scattering study of spin-density-wave-induced anisotropic
  electronic properties in {A}{F}e$_{2}${A}s$_{2}$ ({A}={C}a, {E}u)},}\ }\href
  {\doibase 10.1103/PhysRevB.93.205106} {\bibfield  {journal} {\bibinfo
  {journal} {Phys. Rev. B}\ }\textbf {\bibinfo {volume} {93}},\ \bibinfo
  {pages} {205106} (\bibinfo {year} {2016}{\natexlab{a}})}\BibitemShut
  {NoStop}%
\bibitem [{\citenamefont {Kuroki}\ \emph
  {et~al.}(2009{\natexlab{a}})\citenamefont {Kuroki}, \citenamefont {Usui},
  \citenamefont {Onari}, \citenamefont {Arita},\ and\ \citenamefont
  {Aoki}}]{Kuroki2009PRB}%
  \BibitemOpen
  \bibfield  {author} {\bibinfo {author} {\bibfnamefont {K.}~\bibnamefont
  {Kuroki}}, \bibinfo {author} {\bibfnamefont {H.}~\bibnamefont {Usui}},
  \bibinfo {author} {\bibfnamefont {S.}~\bibnamefont {Onari}}, \bibinfo
  {author} {\bibfnamefont {R.}~\bibnamefont {Arita}}, \ and\ \bibinfo {author}
  {\bibfnamefont {H.}~\bibnamefont {Aoki}},\ }\bibfield  {title} {\enquote
  {\bibinfo {title} {Pnictogen height as a possible switch between
  high-${T}_{c}$ nodeless and low-${T}_{c}$ nodal pairings in the iron-based
  superconductors},}\ }\href {\doibase 10.1103/PhysRevB.79.224511} {\bibfield
  {journal} {\bibinfo  {journal} {Phys. Rev. B}\ }\textbf {\bibinfo {volume}
  {79}},\ \bibinfo {pages} {224511} (\bibinfo {year}
  {2009}{\natexlab{a}})}\BibitemShut {NoStop}%
\bibitem [{\citenamefont {Lee}\ \emph {et~al.}(2014)\citenamefont {Lee},
  \citenamefont {Yun},\ and\ \citenamefont {Hong}}]{Lee2014NJP}%
  \BibitemOpen
  \bibfield  {author} {\bibinfo {author} {\bibfnamefont {J.~D.}\ \bibnamefont
  {Lee}}, \bibinfo {author} {\bibfnamefont {W.~S.}\ \bibnamefont {Yun}}, \ and\
  \bibinfo {author} {\bibfnamefont {S.~C.}\ \bibnamefont {Hong}},\ }\bibfield
  {title} {\enquote {\bibinfo {title} {Ultrafast above-transition-temperature
  resurrection of spin density wave driven by coherent phonon generation in
  {B}a{F}e$_{2}${A}s$_{2}$},}\ }\href
  {http://stacks.iop.org/1367-2630/16/i=4/a=043010} {\bibfield  {journal}
  {\bibinfo  {journal} {New J. Phys.}\ }\textbf {\bibinfo {volume} {16}},\
  \bibinfo {pages} {043010} (\bibinfo {year} {2014})}\BibitemShut {NoStop}%
\bibitem [{\citenamefont {Bal\'edent}\ \emph {et~al.}(2015)\citenamefont
  {Bal\'edent}, \citenamefont {Rullier-Albenque}, \citenamefont {Colson},
  \citenamefont {Ablett},\ and\ \citenamefont {Rueff}}]{Baledent2015PRL}%
  \BibitemOpen
  \bibfield  {author} {\bibinfo {author} {\bibfnamefont {V.}~\bibnamefont
  {Bal\'edent}}, \bibinfo {author} {\bibfnamefont {F.}~\bibnamefont
  {Rullier-Albenque}}, \bibinfo {author} {\bibfnamefont {D.}~\bibnamefont
  {Colson}}, \bibinfo {author} {\bibfnamefont {J.~M.}\ \bibnamefont {Ablett}},
  \ and\ \bibinfo {author} {\bibfnamefont {J.-P.}\ \bibnamefont {Rueff}},\
  }\bibfield  {title} {\enquote {\bibinfo {title} {Electronic properties of
  {B}a{F}e$_2${A}s$_2$ upon doping and pressure: The prominent role of the {A}s
  $p$ orbitals},}\ }\href {\doibase 10.1103/PhysRevLett.114.177001} {\bibfield
  {journal} {\bibinfo  {journal} {Phys. Rev. Lett.}\ }\textbf {\bibinfo
  {volume} {114}},\ \bibinfo {pages} {177001} (\bibinfo {year}
  {2015})}\BibitemShut {NoStop}%
\bibitem [{\citenamefont {Vildosola}\ \emph {et~al.}(2008)\citenamefont
  {Vildosola}, \citenamefont {Pourovskii}, \citenamefont {Arita}, \citenamefont
  {Biermann},\ and\ \citenamefont {Georges}}]{Vildosola2008PRB}%
  \BibitemOpen
  \bibfield  {author} {\bibinfo {author} {\bibfnamefont {V.}~\bibnamefont
  {Vildosola}}, \bibinfo {author} {\bibfnamefont {L.}~\bibnamefont
  {Pourovskii}}, \bibinfo {author} {\bibfnamefont {R.}~\bibnamefont {Arita}},
  \bibinfo {author} {\bibfnamefont {S.}~\bibnamefont {Biermann}}, \ and\
  \bibinfo {author} {\bibfnamefont {A.}~\bibnamefont {Georges}},\ }\bibfield
  {title} {\enquote {\bibinfo {title} {Bandwidth and {F}ermi surface of iron
  oxypnictides: Covalency and sensitivity to structural changes},}\ }\href
  {\doibase 10.1103/PhysRevB.78.064518} {\bibfield  {journal} {\bibinfo
  {journal} {Phys. Rev. B}\ }\textbf {\bibinfo {volume} {78}},\ \bibinfo
  {pages} {064518} (\bibinfo {year} {2008})}\BibitemShut {NoStop}%
\bibitem [{\citenamefont {Calder\'on}\ \emph {et~al.}(2009)\citenamefont
  {Calder\'on}, \citenamefont {Valenzuela},\ and\ \citenamefont
  {Bascones}}]{Bascones2009PRB}%
  \BibitemOpen
  \bibfield  {author} {\bibinfo {author} {\bibfnamefont {M.~J.}\ \bibnamefont
  {Calder\'on}}, \bibinfo {author} {\bibfnamefont {B.}~\bibnamefont
  {Valenzuela}}, \ and\ \bibinfo {author} {\bibfnamefont {E.}~\bibnamefont
  {Bascones}},\ }\bibfield  {title} {\enquote {\bibinfo {title} {Tight-binding
  model for iron pnictides},}\ }\href {\doibase 10.1103/PhysRevB.80.094531}
  {\bibfield  {journal} {\bibinfo  {journal} {Phys. Rev. B}\ }\textbf {\bibinfo
  {volume} {80}},\ \bibinfo {pages} {094531} (\bibinfo {year}
  {2009})}\BibitemShut {NoStop}%
\bibitem [{\citenamefont {Yin}\ \emph {et~al.}(2008)\citenamefont {Yin},
  \citenamefont {Leb\`egue}, \citenamefont {Han}, \citenamefont {Neal},
  \citenamefont {Savrasov},\ and\ \citenamefont {Pickett}}]{YinPRL2008}%
  \BibitemOpen
  \bibfield  {author} {\bibinfo {author} {\bibfnamefont {Z.~P.}\ \bibnamefont
  {Yin}}, \bibinfo {author} {\bibfnamefont {S.}~\bibnamefont {Leb\`egue}},
  \bibinfo {author} {\bibfnamefont {M.~J.}\ \bibnamefont {Han}}, \bibinfo
  {author} {\bibfnamefont {B.~P.}\ \bibnamefont {Neal}}, \bibinfo {author}
  {\bibfnamefont {S.~Y.}\ \bibnamefont {Savrasov}}, \ and\ \bibinfo {author}
  {\bibfnamefont {W.~E.}\ \bibnamefont {Pickett}},\ }\bibfield  {title}
  {\enquote {\bibinfo {title} {Electron-hole symmetry and magnetic coupling in
  antiferromagnetic {L}a{F}e{A}s{O}},}\ }\href {\doibase
  10.1103/PhysRevLett.101.047001} {\bibfield  {journal} {\bibinfo  {journal}
  {Phys. Rev. Lett.}\ }\textbf {\bibinfo {volume} {101}},\ \bibinfo {pages}
  {047001} (\bibinfo {year} {2008})}\BibitemShut {NoStop}%
\bibitem [{\citenamefont {Yndurain}(2011)}]{Yndurain2011EPL}%
  \BibitemOpen
  \bibfield  {author} {\bibinfo {author} {\bibfnamefont {F.}~\bibnamefont
  {Yndurain}},\ }\bibfield  {title} {\enquote {\bibinfo {title} {Coupling of
  magnetic moments with phonons and electron-phonon interaction in
  {L}a{F}e{A}s{O}$_{1-x}$ {F}$_x$},}\ }\href
  {http://stacks.iop.org/0295-5075/94/i=3/a=37001} {\bibfield  {journal}
  {\bibinfo  {journal} {EPL}\ }\textbf {\bibinfo {volume} {94}},\ \bibinfo
  {pages} {37001} (\bibinfo {year} {2011})}\BibitemShut {NoStop}%
\bibitem [{\citenamefont {de~la Cruz}\ \emph {et~al.}(2010)\citenamefont {de~la
  Cruz}, \citenamefont {Hu}, \citenamefont {Li}, \citenamefont {Huang},
  \citenamefont {Lynn}, \citenamefont {Green}, \citenamefont {Chen},
  \citenamefont {Wang}, \citenamefont {Mook}, \citenamefont {Si},\ and\
  \citenamefont {Dai}}]{Cruz2010PRL}%
  \BibitemOpen
  \bibfield  {author} {\bibinfo {author} {\bibfnamefont {C.}~\bibnamefont
  {de~la Cruz}}, \bibinfo {author} {\bibfnamefont {W.~Z.}\ \bibnamefont {Hu}},
  \bibinfo {author} {\bibfnamefont {S.~L}\ \bibnamefont {Li}}, \bibinfo
  {author} {\bibfnamefont {Q.}~\bibnamefont {Huang}}, \bibinfo {author}
  {\bibfnamefont {J.~W.}\ \bibnamefont {Lynn}}, \bibinfo {author}
  {\bibfnamefont {M.~A.}\ \bibnamefont {Green}}, \bibinfo {author}
  {\bibfnamefont {G.~F.}\ \bibnamefont {Chen}}, \bibinfo {author}
  {\bibfnamefont {N.~L.}\ \bibnamefont {Wang}}, \bibinfo {author}
  {\bibfnamefont {H.~A.}\ \bibnamefont {Mook}}, \bibinfo {author}
  {\bibfnamefont {Q.~M}\ \bibnamefont {Si}}, \ and\ \bibinfo {author}
  {\bibfnamefont {P.~C}\ \bibnamefont {Dai}},\ }\bibfield  {title} {\enquote
  {\bibinfo {title} {Lattice distortion and magnetic quantum phase transition
  in {C}e{F}e{A}s$_{1-x}${P}$_{x}${O}},}\ }\href {\doibase
  10.1103/PhysRevLett.104.017204} {\bibfield  {journal} {\bibinfo  {journal}
  {Phys. Rev. Lett.}\ }\textbf {\bibinfo {volume} {104}},\ \bibinfo {pages}
  {017204} (\bibinfo {year} {2010})}\BibitemShut {NoStop}%
\bibitem [{\citenamefont {Zhang}\ \emph
  {et~al.}(2014{\natexlab{a}})\citenamefont {Zhang}, \citenamefont {Harriger},
  \citenamefont {Yin}, \citenamefont {Lv}, \citenamefont {Wang}, \citenamefont
  {Tan}, \citenamefont {Song}, \citenamefont {Abernathy}, \citenamefont {Tian},
  \citenamefont {Egami}, \citenamefont {Haule}, \citenamefont {Kotliar},\ and\
  \citenamefont {Dai}}]{Zhang2014PRL}%
  \BibitemOpen
  \bibfield  {author} {\bibinfo {author} {\bibfnamefont {C.~L.}\ \bibnamefont
  {Zhang}}, \bibinfo {author} {\bibfnamefont {L.~W.}\ \bibnamefont {Harriger}},
  \bibinfo {author} {\bibfnamefont {Z.~P.}\ \bibnamefont {Yin}}, \bibinfo
  {author} {\bibfnamefont {W.~C.}\ \bibnamefont {Lv}}, \bibinfo {author}
  {\bibfnamefont {M.~Y.}\ \bibnamefont {Wang}}, \bibinfo {author}
  {\bibfnamefont {G.~T.}\ \bibnamefont {Tan}}, \bibinfo {author} {\bibfnamefont
  {Y.}~\bibnamefont {Song}}, \bibinfo {author} {\bibfnamefont {D.~L.}\
  \bibnamefont {Abernathy}}, \bibinfo {author} {\bibfnamefont {W.}~\bibnamefont
  {Tian}}, \bibinfo {author} {\bibfnamefont {T.}~\bibnamefont {Egami}},
  \bibinfo {author} {\bibfnamefont {K.}~\bibnamefont {Haule}}, \bibinfo
  {author} {\bibfnamefont {G.}~\bibnamefont {Kotliar}}, \ and\ \bibinfo
  {author} {\bibfnamefont {P.~C.}\ \bibnamefont {Dai}},\ }\bibfield  {title}
  {\enquote {\bibinfo {title} {Effect of pnictogen height on spin waves in iron
  pnictides},}\ }\href {\doibase 10.1103/PhysRevLett.112.217202} {\bibfield
  {journal} {\bibinfo  {journal} {Phys. Rev. Lett.}\ }\textbf {\bibinfo
  {volume} {112}},\ \bibinfo {pages} {217202} (\bibinfo {year}
  {2014}{\natexlab{a}})}\BibitemShut {NoStop}%
\bibitem [{\citenamefont {Lee}\ \emph {et~al.}(2008)\citenamefont {Lee},
  \citenamefont {Iyo}, \citenamefont {Eisaki}, \citenamefont {Kito},
  \citenamefont {Fernandez-Diaz}, \citenamefont {Ito}, \citenamefont {Kihou},
  \citenamefont {Matsuhata}, \citenamefont {Braden},\ and\ \citenamefont
  {Yamada}}]{Lee2008JPSJ}%
  \BibitemOpen
  \bibfield  {author} {\bibinfo {author} {\bibfnamefont {C.~H}\ \bibnamefont
  {Lee}}, \bibinfo {author} {\bibfnamefont {A.}~\bibnamefont {Iyo}}, \bibinfo
  {author} {\bibfnamefont {H.}~\bibnamefont {Eisaki}}, \bibinfo {author}
  {\bibfnamefont {H.}~\bibnamefont {Kito}}, \bibinfo {author} {\bibfnamefont
  {M.~T.}\ \bibnamefont {Fernandez-Diaz}}, \bibinfo {author} {\bibfnamefont
  {T.}~\bibnamefont {Ito}}, \bibinfo {author} {\bibfnamefont {K.}~\bibnamefont
  {Kihou}}, \bibinfo {author} {\bibfnamefont {H.}~\bibnamefont {Matsuhata}},
  \bibinfo {author} {\bibfnamefont {M.}~\bibnamefont {Braden}}, \ and\ \bibinfo
  {author} {\bibfnamefont {K.}~\bibnamefont {Yamada}},\ }\bibfield  {title}
  {\enquote {\bibinfo {title} {Effect of structural parameters on
  superconductivity in fluorine-free {L}n{F}e{A}s{O}$_{1-y}$ ({L}n = {L}a,
  {N}d)},}\ }\href {\doibase 10.1143/JPSJ.77.083704} {\bibfield  {journal}
  {\bibinfo  {journal} {J. Phys. Soc. Jpn.}\ }\textbf {\bibinfo {volume}
  {77}},\ \bibinfo {pages} {083704} (\bibinfo {year} {2008})}\BibitemShut
  {NoStop}%
\bibitem [{\citenamefont {Zhao}\ \emph {et~al.}(2008)\citenamefont {Zhao},
  \citenamefont {Huang}, \citenamefont {de~la Cruz}, \citenamefont {Li},
  \citenamefont {Lynn}, \citenamefont {Chen}, \citenamefont {Green},
  \citenamefont {Chen}, \citenamefont {Li}, \citenamefont {Li}, \citenamefont
  {Luo}, \citenamefont {Wang},\ and\ \citenamefont {Dai}}]{Zhao2008}%
  \BibitemOpen
  \bibfield  {author} {\bibinfo {author} {\bibfnamefont {J.}~\bibnamefont
  {Zhao}}, \bibinfo {author} {\bibfnamefont {Q.}~\bibnamefont {Huang}},
  \bibinfo {author} {\bibfnamefont {C.}~\bibnamefont {de~la Cruz}}, \bibinfo
  {author} {\bibfnamefont {S.~L.}\ \bibnamefont {Li}}, \bibinfo {author}
  {\bibfnamefont {J.~W.}\ \bibnamefont {Lynn}}, \bibinfo {author}
  {\bibfnamefont {Y.}~\bibnamefont {Chen}}, \bibinfo {author} {\bibfnamefont
  {M.~A.}\ \bibnamefont {Green}}, \bibinfo {author} {\bibfnamefont {G.~F.}\
  \bibnamefont {Chen}}, \bibinfo {author} {\bibfnamefont {G.}~\bibnamefont
  {Li}}, \bibinfo {author} {\bibfnamefont {Z.}~\bibnamefont {Li}}, \bibinfo
  {author} {\bibfnamefont {J.~L.}\ \bibnamefont {Luo}}, \bibinfo {author}
  {\bibfnamefont {N.~L.}\ \bibnamefont {Wang}}, \ and\ \bibinfo {author}
  {\bibfnamefont {P.~C.}\ \bibnamefont {Dai}},\ }\bibfield  {title} {\enquote
  {\bibinfo {title} {Structural and magnetic phase diagram of
  {C}e{F}e{A}s{O}$_{1- x}${F}$_x$ and its relation to high-temperature
  superconductivity},}\ }\href {\doibase 10.1038/nmat2315} {\bibfield
  {journal} {\bibinfo  {journal} {Nat. Mater.}\ }\textbf {\bibinfo {volume}
  {7}},\ \bibinfo {pages} {953--959} (\bibinfo {year} {2008})}\BibitemShut
  {NoStop}%
\bibitem [{\citenamefont {Kuroki}\ \emph
  {et~al.}(2009{\natexlab{b}})\citenamefont {Kuroki}, \citenamefont {Usui},
  \citenamefont {Onari}, \citenamefont {Arita},\ and\ \citenamefont
  {Aoki}}]{KurokiPRB2009}%
  \BibitemOpen
  \bibfield  {author} {\bibinfo {author} {\bibfnamefont {K.}~\bibnamefont
  {Kuroki}}, \bibinfo {author} {\bibfnamefont {H.}~\bibnamefont {Usui}},
  \bibinfo {author} {\bibfnamefont {S.}~\bibnamefont {Onari}}, \bibinfo
  {author} {\bibfnamefont {R.}~\bibnamefont {Arita}}, \ and\ \bibinfo {author}
  {\bibfnamefont {H.}~\bibnamefont {Aoki}},\ }\bibfield  {title} {\enquote
  {\bibinfo {title} {Pnictogen height as a possible switch between
  high-${T}_{c}$ nodeless and low-${T}_{c}$ nodal pairings in the iron-based
  superconductors},}\ }\href {\doibase 10.1103/PhysRevB.79.224511} {\bibfield
  {journal} {\bibinfo  {journal} {Phys. Rev. B}\ }\textbf {\bibinfo {volume}
  {79}},\ \bibinfo {pages} {224511} (\bibinfo {year}
  {2009}{\natexlab{b}})}\BibitemShut {NoStop}%
\bibitem [{\citenamefont {Garbarino}\ \emph {et~al.}(2011)\citenamefont
  {Garbarino}, \citenamefont {Weht}, \citenamefont {A.Sow}, \citenamefont
  {Lacroix}, \citenamefont {Sulpice}, \citenamefont {Mezouar}, \citenamefont
  {Zhu}, \citenamefont {Han}, \citenamefont {Wen},\ and\ \citenamefont
  {Núñez-Regueiro}}]{Garbarino2011EPL}%
  \BibitemOpen
  \bibfield  {author} {\bibinfo {author} {\bibfnamefont {G.}~\bibnamefont
  {Garbarino}}, \bibinfo {author} {\bibfnamefont {R.}~\bibnamefont {Weht}},
  \bibinfo {author} {\bibnamefont {A.Sow}}, \bibinfo {author} {\bibfnamefont
  {C.}~\bibnamefont {Lacroix}}, \bibinfo {author} {\bibfnamefont
  {A.}~\bibnamefont {Sulpice}}, \bibinfo {author} {\bibfnamefont
  {M.}~\bibnamefont {Mezouar}}, \bibinfo {author} {\bibfnamefont
  {X.}~\bibnamefont {Zhu}}, \bibinfo {author} {\bibfnamefont {F.}~\bibnamefont
  {Han}}, \bibinfo {author} {\bibfnamefont {H.~H.}\ \bibnamefont {Wen}}, \ and\
  \bibinfo {author} {\bibfnamefont {M.}~\bibnamefont {Núñez-Regueiro}},\
  }\bibfield  {title} {\enquote {\bibinfo {title} {Direct observation of the
  influence of the {F}e{A}s$_4$ tetrahedron on superconductivity and
  antiferromagnetic correlations in {S}r$_2${V}{O}$_3${F}e{A}s},}\ }\href
  {http://stacks.iop.org/0295-5075/96/i=5/a=57002} {\bibfield  {journal}
  {\bibinfo  {journal} {EPL}\ }\textbf {\bibinfo {volume} {96}},\ \bibinfo
  {pages} {57002} (\bibinfo {year} {2011})}\BibitemShut {NoStop}%
\bibitem [{\citenamefont {Garc\'{\i}a-Mart\'{\i}nez}\ \emph
  {et~al.}(2013)\citenamefont {Garc\'{\i}a-Mart\'{\i}nez}, \citenamefont
  {Valenzuela}, \citenamefont {Ciuchi}, \citenamefont {Cappelluti},
  \citenamefont {Calder\'on},\ and\ \citenamefont
  {Bascones}}]{Bascones2013PRB}%
  \BibitemOpen
  \bibfield  {author} {\bibinfo {author} {\bibfnamefont {N.~A.}\ \bibnamefont
  {Garc\'{\i}a-Mart\'{\i}nez}}, \bibinfo {author} {\bibfnamefont
  {B.}~\bibnamefont {Valenzuela}}, \bibinfo {author} {\bibfnamefont
  {S.}~\bibnamefont {Ciuchi}}, \bibinfo {author} {\bibfnamefont
  {E.}~\bibnamefont {Cappelluti}}, \bibinfo {author} {\bibfnamefont {M.~J.}\
  \bibnamefont {Calder\'on}}, \ and\ \bibinfo {author} {\bibfnamefont
  {E.}~\bibnamefont {Bascones}},\ }\bibfield  {title} {\enquote {\bibinfo
  {title} {Coupling of the {A}s ${A}_{1g}$ phonon to magnetism in iron
  pnictides},}\ }\href {\doibase 10.1103/PhysRevB.88.165106} {\bibfield
  {journal} {\bibinfo  {journal} {Phys. Rev. B}\ }\textbf {\bibinfo {volume}
  {88}},\ \bibinfo {pages} {165106} (\bibinfo {year} {2013})}\BibitemShut
  {NoStop}%
\bibitem [{\citenamefont {Mansart}\ \emph {et~al.}(2009)\citenamefont
  {Mansart}, \citenamefont {Boschetto}, \citenamefont {Savoia}, \citenamefont
  {Rullier-Albenque}, \citenamefont {Forget}, \citenamefont {Colson},
  \citenamefont {Rousse},\ and\ \citenamefont {Marsi}}]{Mansart2009PRB}%
  \BibitemOpen
  \bibfield  {author} {\bibinfo {author} {\bibfnamefont {B.}~\bibnamefont
  {Mansart}}, \bibinfo {author} {\bibfnamefont {D.}~\bibnamefont {Boschetto}},
  \bibinfo {author} {\bibfnamefont {A.}~\bibnamefont {Savoia}}, \bibinfo
  {author} {\bibfnamefont {F.}~\bibnamefont {Rullier-Albenque}}, \bibinfo
  {author} {\bibfnamefont {A.}~\bibnamefont {Forget}}, \bibinfo {author}
  {\bibfnamefont {D.}~\bibnamefont {Colson}}, \bibinfo {author} {\bibfnamefont
  {A.}~\bibnamefont {Rousse}}, \ and\ \bibinfo {author} {\bibfnamefont
  {M.}~\bibnamefont {Marsi}},\ }\bibfield  {title} {\enquote {\bibinfo {title}
  {Observation of a coherent optical phonon in the iron pnictide superconductor
  {B}a({F}e$_{1-x}${C}o$_{x}$)$_{2}${A}s$_{2}$ ($x=0.06$ and $0.08$)},}\ }\href
  {\doibase 10.1103/PhysRevB.80.172504} {\bibfield  {journal} {\bibinfo
  {journal} {Phys. Rev. B}\ }\textbf {\bibinfo {volume} {80}},\ \bibinfo
  {pages} {172504} (\bibinfo {year} {2009})}\BibitemShut {NoStop}%
\bibitem [{\citenamefont {Kim}\ \emph {et~al.}(2012)\citenamefont {Kim},
  \citenamefont {Pashkin}, \citenamefont {Sch{\"a}fer}, \citenamefont {Beyer},
  \citenamefont {Porer}, \citenamefont {Wolf}, \citenamefont {Bernhard},
  \citenamefont {Demsar}, \citenamefont {Huber},\ and\ \citenamefont
  {Leitenstorfer}}]{Kim2012NatMat}%
  \BibitemOpen
  \bibfield  {author} {\bibinfo {author} {\bibfnamefont {K.~W.}\ \bibnamefont
  {Kim}}, \bibinfo {author} {\bibfnamefont {A.}~\bibnamefont {Pashkin}},
  \bibinfo {author} {\bibfnamefont {H.}~\bibnamefont {Sch{\"a}fer}}, \bibinfo
  {author} {\bibfnamefont {M.}~\bibnamefont {Beyer}}, \bibinfo {author}
  {\bibfnamefont {M.}~\bibnamefont {Porer}}, \bibinfo {author} {\bibfnamefont
  {T.}~\bibnamefont {Wolf}}, \bibinfo {author} {\bibfnamefont {C.}~\bibnamefont
  {Bernhard}}, \bibinfo {author} {\bibfnamefont {J.}~\bibnamefont {Demsar}},
  \bibinfo {author} {\bibfnamefont {R.}~\bibnamefont {Huber}}, \ and\ \bibinfo
  {author} {\bibfnamefont {A.}~\bibnamefont {Leitenstorfer}},\ }\bibfield
  {title} {\enquote {\bibinfo {title} {Ultrafast transient generation of
  spin-density-wave order in the normal state of {B}a{F}e$_2${A}s$_2$ driven by
  coherent lattice vibrations},}\ }\href {\doibase 10.1038/nmat3294} {\bibfield
   {journal} {\bibinfo  {journal} {Nat. Mater.}\ }\textbf {\bibinfo {volume}
  {11}},\ \bibinfo {pages} {497--501} (\bibinfo {year} {2012})}\BibitemShut
  {NoStop}%
\bibitem [{\citenamefont {Avigo}\ \emph {et~al.}(2013)\citenamefont {Avigo},
  \citenamefont {Cortés}, \citenamefont {Rettig}, \citenamefont
  {Thirupathaiah}, \citenamefont {Jeevan}, \citenamefont {Gegenwart},
  \citenamefont {Wolf}, \citenamefont {Ligges}, \citenamefont {Wolf},
  \citenamefont {Fink},\ and\ \citenamefont {Bovensiepen}}]{Avigo2013JPCM}%
  \BibitemOpen
  \bibfield  {author} {\bibinfo {author} {\bibfnamefont {I.}~\bibnamefont
  {Avigo}}, \bibinfo {author} {\bibfnamefont {R.}~\bibnamefont {Cortés}},
  \bibinfo {author} {\bibfnamefont {L.}~\bibnamefont {Rettig}}, \bibinfo
  {author} {\bibfnamefont {S.}~\bibnamefont {Thirupathaiah}}, \bibinfo {author}
  {\bibfnamefont {H.~S.}\ \bibnamefont {Jeevan}}, \bibinfo {author}
  {\bibfnamefont {P.}~\bibnamefont {Gegenwart}}, \bibinfo {author}
  {\bibfnamefont {T.}~\bibnamefont {Wolf}}, \bibinfo {author} {\bibfnamefont
  {M.}~\bibnamefont {Ligges}}, \bibinfo {author} {\bibfnamefont
  {M.}~\bibnamefont {Wolf}}, \bibinfo {author} {\bibfnamefont {J.}~\bibnamefont
  {Fink}}, \ and\ \bibinfo {author} {\bibfnamefont {U.}~\bibnamefont
  {Bovensiepen}},\ }\bibfield  {title} {\enquote {\bibinfo {title} {Coherent
  excitations and electron–phonon coupling in {B}a/{E}u{F}e$_2${A}s$_2$
  compounds investigated by femtosecond time- and angle-resolved photoemission
  spectroscopy},}\ }\href {http://stacks.iop.org/0953-8984/25/i=9/a=094003}
  {\bibfield  {journal} {\bibinfo  {journal} {J. Phys.: Condens. Matter}\
  }\textbf {\bibinfo {volume} {25}},\ \bibinfo {pages} {094003} (\bibinfo
  {year} {2013})}\BibitemShut {NoStop}%
\bibitem [{\citenamefont {Yang}\ \emph {et~al.}(2014)\citenamefont {Yang},
  \citenamefont {Rohde}, \citenamefont {Rohwer}, \citenamefont {Stange},
  \citenamefont {Hanff}, \citenamefont {Sohrt}, \citenamefont {Rettig},
  \citenamefont {Cort\'es}, \citenamefont {Chen}, \citenamefont {Feng},
  \citenamefont {Wolf}, \citenamefont {Kamble}, \citenamefont {Eremin},
  \citenamefont {Popmintchev}, \citenamefont {Murnane}, \citenamefont
  {Kapteyn}, \citenamefont {Kipp}, \citenamefont {Fink}, \citenamefont {Bauer},
  \citenamefont {Bovensiepen},\ and\ \citenamefont {Rossnagel}}]{Yang2014PRL}%
  \BibitemOpen
  \bibfield  {author} {\bibinfo {author} {\bibfnamefont {L.~X.}\ \bibnamefont
  {Yang}}, \bibinfo {author} {\bibfnamefont {G.}~\bibnamefont {Rohde}},
  \bibinfo {author} {\bibfnamefont {T.}~\bibnamefont {Rohwer}}, \bibinfo
  {author} {\bibfnamefont {A.}~\bibnamefont {Stange}}, \bibinfo {author}
  {\bibfnamefont {K.}~\bibnamefont {Hanff}}, \bibinfo {author} {\bibfnamefont
  {C.}~\bibnamefont {Sohrt}}, \bibinfo {author} {\bibfnamefont
  {L.}~\bibnamefont {Rettig}}, \bibinfo {author} {\bibfnamefont
  {R.}~\bibnamefont {Cort\'es}}, \bibinfo {author} {\bibfnamefont
  {F.}~\bibnamefont {Chen}}, \bibinfo {author} {\bibfnamefont {D.~L.}\
  \bibnamefont {Feng}}, \bibinfo {author} {\bibfnamefont {T.}~\bibnamefont
  {Wolf}}, \bibinfo {author} {\bibfnamefont {B.}~\bibnamefont {Kamble}},
  \bibinfo {author} {\bibfnamefont {I.}~\bibnamefont {Eremin}}, \bibinfo
  {author} {\bibfnamefont {T.}~\bibnamefont {Popmintchev}}, \bibinfo {author}
  {\bibfnamefont {M.~M.}\ \bibnamefont {Murnane}}, \bibinfo {author}
  {\bibfnamefont {H.~C.}\ \bibnamefont {Kapteyn}}, \bibinfo {author}
  {\bibfnamefont {L.}~\bibnamefont {Kipp}}, \bibinfo {author} {\bibfnamefont
  {J.}~\bibnamefont {Fink}}, \bibinfo {author} {\bibfnamefont {M.}~\bibnamefont
  {Bauer}}, \bibinfo {author} {\bibfnamefont {U.}~\bibnamefont {Bovensiepen}},
  \ and\ \bibinfo {author} {\bibfnamefont {K.}~\bibnamefont {Rossnagel}},\
  }\bibfield  {title} {\enquote {\bibinfo {title} {Ultrafast modulation of the
  chemical potential in {B}a{F}e$_2${A}s$_2$ by coherent phonons},}\ }\href
  {\doibase 10.1103/PhysRevLett.112.207001} {\bibfield  {journal} {\bibinfo
  {journal} {Phys. Rev. Lett.}\ }\textbf {\bibinfo {volume} {112}},\ \bibinfo
  {pages} {207001} (\bibinfo {year} {2014})}\BibitemShut {NoStop}%
\bibitem [{\citenamefont {Gerber}\ \emph {et~al.}(2015)\citenamefont {Gerber},
  \citenamefont {Kim}, \citenamefont {Zhang}, \citenamefont {Zhu},
  \citenamefont {Plonka}, \citenamefont {Yi}, \citenamefont {Dakovski},
  \citenamefont {Leuenberger}, \citenamefont {Kirchmann}, \citenamefont
  {Moore}, \citenamefont {Chollet}, \citenamefont {Glownia}, \citenamefont
  {Feng}, \citenamefont {Lee}, \citenamefont {Mehta}, \citenamefont {Kemper},
  \citenamefont {Wolf}, \citenamefont {Chuang}, \citenamefont {Hussain},
  \citenamefont {Kao}, \citenamefont {Moritz}, \citenamefont {Shen},
  \citenamefont {Devereaux},\ and\ \citenamefont {Lee}}]{Gerber2015NatCom}%
  \BibitemOpen
  \bibfield  {author} {\bibinfo {author} {\bibfnamefont {S.}~\bibnamefont
  {Gerber}}, \bibinfo {author} {\bibfnamefont {K.~W.}\ \bibnamefont {Kim}},
  \bibinfo {author} {\bibfnamefont {Y.}~\bibnamefont {Zhang}}, \bibinfo
  {author} {\bibfnamefont {D.}~\bibnamefont {Zhu}}, \bibinfo {author}
  {\bibfnamefont {N.}~\bibnamefont {Plonka}}, \bibinfo {author} {\bibfnamefont
  {M.}~\bibnamefont {Yi}}, \bibinfo {author} {\bibfnamefont {G.~L.}\
  \bibnamefont {Dakovski}}, \bibinfo {author} {\bibfnamefont {D.}~\bibnamefont
  {Leuenberger}}, \bibinfo {author} {\bibfnamefont {P.~S.}\ \bibnamefont
  {Kirchmann}}, \bibinfo {author} {\bibfnamefont {R.~G.}\ \bibnamefont
  {Moore}}, \bibinfo {author} {\bibfnamefont {M.}~\bibnamefont {Chollet}},
  \bibinfo {author} {\bibfnamefont {J.~M.}\ \bibnamefont {Glownia}}, \bibinfo
  {author} {\bibfnamefont {Y.}~\bibnamefont {Feng}}, \bibinfo {author}
  {\bibfnamefont {J.-S.}\ \bibnamefont {Lee}}, \bibinfo {author} {\bibfnamefont
  {A.}~\bibnamefont {Mehta}}, \bibinfo {author} {\bibfnamefont {A.~F.}\
  \bibnamefont {Kemper}}, \bibinfo {author} {\bibfnamefont {T.}~\bibnamefont
  {Wolf}}, \bibinfo {author} {\bibfnamefont {Y.-D.}\ \bibnamefont {Chuang}},
  \bibinfo {author} {\bibfnamefont {Z.}~\bibnamefont {Hussain}}, \bibinfo
  {author} {\bibfnamefont {C.-C.}\ \bibnamefont {Kao}}, \bibinfo {author}
  {\bibfnamefont {B.}~\bibnamefont {Moritz}}, \bibinfo {author} {\bibfnamefont
  {Z.-X.}\ \bibnamefont {Shen}}, \bibinfo {author} {\bibfnamefont {T.~P.}\
  \bibnamefont {Devereaux}}, \ and\ \bibinfo {author} {\bibfnamefont {W.-S.}\
  \bibnamefont {Lee}},\ }\bibfield  {title} {\enquote {\bibinfo {title} {Direct
  characterization of photoinduced lattice dynamics in {B}a{F}e$_2${A}s$_2$},}\
  }\href {https://www.nature.com/articles/ncomms8377} {\bibfield  {journal}
  {\bibinfo  {journal} {Nat. Commun.}\ }\textbf {\bibinfo {volume} {6}},\
  \bibinfo {pages} {7377} (\bibinfo {year} {2015})}\BibitemShut {NoStop}%
\bibitem [{\citenamefont {Rettig}\ \emph {et~al.}(2015)\citenamefont {Rettig},
  \citenamefont {Mariager}, \citenamefont {Ferrer}, \citenamefont {Gr\"ubel},
  \citenamefont {Johnson}, \citenamefont {Rittmann}, \citenamefont {Wolf},
  \citenamefont {Johnson}, \citenamefont {Ingold}, \citenamefont {Beaud},\ and\
  \citenamefont {Staub}}]{Rettig2015PhysRevLett114}%
  \BibitemOpen
  \bibfield  {author} {\bibinfo {author} {\bibfnamefont {L.}~\bibnamefont
  {Rettig}}, \bibinfo {author} {\bibfnamefont {S.~O.}\ \bibnamefont
  {Mariager}}, \bibinfo {author} {\bibfnamefont {A.}~\bibnamefont {Ferrer}},
  \bibinfo {author} {\bibfnamefont {S.}~\bibnamefont {Gr\"ubel}}, \bibinfo
  {author} {\bibfnamefont {J.~A.}\ \bibnamefont {Johnson}}, \bibinfo {author}
  {\bibfnamefont {J.}~\bibnamefont {Rittmann}}, \bibinfo {author}
  {\bibfnamefont {T.}~\bibnamefont {Wolf}}, \bibinfo {author} {\bibfnamefont
  {S.~L.}\ \bibnamefont {Johnson}}, \bibinfo {author} {\bibfnamefont
  {G.}~\bibnamefont {Ingold}}, \bibinfo {author} {\bibfnamefont
  {P.}~\bibnamefont {Beaud}}, \ and\ \bibinfo {author} {\bibfnamefont
  {U.}~\bibnamefont {Staub}},\ }\bibfield  {title} {\enquote {\bibinfo {title}
  {Ultrafast structural dynamics of the {F}e-pnictide parent compound
  {B}a{F}e$_2${A}s$_2$},}\ }\href {\doibase 10.1103/PhysRevLett.114.067402}
  {\bibfield  {journal} {\bibinfo  {journal} {Phys. Rev. Lett.}\ }\textbf
  {\bibinfo {volume} {114}},\ \bibinfo {pages} {067402} (\bibinfo {year}
  {2015})}\BibitemShut {NoStop}%
\bibitem [{\citenamefont {Mandal}\ \emph {et~al.}(2014)\citenamefont {Mandal},
  \citenamefont {Cohen},\ and\ \citenamefont {Haule}}]{Mandal2014PRB}%
  \BibitemOpen
  \bibfield  {author} {\bibinfo {author} {\bibfnamefont {S.}~\bibnamefont
  {Mandal}}, \bibinfo {author} {\bibfnamefont {R.~E.}\ \bibnamefont {Cohen}}, \
  and\ \bibinfo {author} {\bibfnamefont {K.}~\bibnamefont {Haule}},\ }\bibfield
   {title} {\enquote {\bibinfo {title} {Strong pressure-dependent
  electron-phonon coupling in {F}e{S}e},}\ }\href {\doibase
  10.1103/PhysRevB.89.220502} {\bibfield  {journal} {\bibinfo  {journal} {Phys.
  Rev. B}\ }\textbf {\bibinfo {volume} {89}},\ \bibinfo {pages} {220502}
  (\bibinfo {year} {2014})}\BibitemShut {NoStop}%
\bibitem [{\citenamefont {Yildirim}(2009)}]{Yildirim2009PhysicaC}%
  \BibitemOpen
  \bibfield  {author} {\bibinfo {author} {\bibfnamefont {T.}~\bibnamefont
  {Yildirim}},\ }\bibfield  {title} {\enquote {\bibinfo {title} {Frustrated
  magnetic interactions, giant magneto–elastic coupling, and magnetic phonons
  in iron pnictides},}\ }\href {\doibase
  http://dx.doi.org/10.1016/j.physc.2009.03.038} {\bibfield  {journal}
  {\bibinfo  {journal} {Physica C}\ }\textbf {\bibinfo {volume} {469}},\
  \bibinfo {pages} {425} (\bibinfo {year} {2009})}\BibitemShut {NoStop}%
\bibitem [{\citenamefont {Boeri}\ \emph
  {et~al.}(2010{\natexlab{a}})\citenamefont {Boeri}, \citenamefont {Calandra},
  \citenamefont {Mazin}, \citenamefont {Dolgov},\ and\ \citenamefont
  {Mauri}}]{BoeriPRB2010}%
  \BibitemOpen
  \bibfield  {author} {\bibinfo {author} {\bibfnamefont {L.}~\bibnamefont
  {Boeri}}, \bibinfo {author} {\bibfnamefont {M.}~\bibnamefont {Calandra}},
  \bibinfo {author} {\bibfnamefont {I.~I.}\ \bibnamefont {Mazin}}, \bibinfo
  {author} {\bibfnamefont {O.~V.}\ \bibnamefont {Dolgov}}, \ and\ \bibinfo
  {author} {\bibfnamefont {F.}~\bibnamefont {Mauri}},\ }\bibfield  {title}
  {\enquote {\bibinfo {title} {Effects of magnetism and doping on the
  electron-phonon coupling in {B}a{F}e$_2${A}s$_{2}$},}\ }\href {\doibase
  10.1103/PhysRevB.82.020506} {\bibfield  {journal} {\bibinfo  {journal} {Phys.
  Rev. B}\ }\textbf {\bibinfo {volume} {82}},\ \bibinfo {pages} {020506}
  (\bibinfo {year} {2010}{\natexlab{a}})}\BibitemShut {NoStop}%
\bibitem [{\citenamefont {Zbiri}\ \emph {et~al.}(2009)\citenamefont {Zbiri},
  \citenamefont {Schober}, \citenamefont {Johnson}, \citenamefont {Rols},
  \citenamefont {Mittal}, \citenamefont {Su}, \citenamefont {Rotter},\ and\
  \citenamefont {Johrendt}}]{Zbiri2009PRB}%
  \BibitemOpen
  \bibfield  {author} {\bibinfo {author} {\bibfnamefont {M.}~\bibnamefont
  {Zbiri}}, \bibinfo {author} {\bibfnamefont {H.}~\bibnamefont {Schober}},
  \bibinfo {author} {\bibfnamefont {M.~R.}\ \bibnamefont {Johnson}}, \bibinfo
  {author} {\bibfnamefont {S.}~\bibnamefont {Rols}}, \bibinfo {author}
  {\bibfnamefont {R.}~\bibnamefont {Mittal}}, \bibinfo {author} {\bibfnamefont
  {Y.~X.}\ \bibnamefont {Su}}, \bibinfo {author} {\bibfnamefont
  {M.}~\bibnamefont {Rotter}}, \ and\ \bibinfo {author} {\bibfnamefont
  {D.}~\bibnamefont {Johrendt}},\ }\bibfield  {title} {\enquote {\bibinfo
  {title} {$ab$ initio lattice dynamics simulations and inelastic neutron
  scattering spectra for studying phonons in {B}a{F}e$_2${A}s$_2$: Effect of
  structural phase transition, structural relaxation, and magnetic ordering},}\
  }\href {\doibase 10.1103/PhysRevB.79.064511} {\bibfield  {journal} {\bibinfo
  {journal} {Phys. Rev. B}\ }\textbf {\bibinfo {volume} {79}},\ \bibinfo
  {pages} {064511} (\bibinfo {year} {2009})}\BibitemShut {NoStop}%
\bibitem [{\citenamefont {Reznik}\ \emph {et~al.}(2009)\citenamefont {Reznik},
  \citenamefont {Lokshin}, \citenamefont {Mitchell}, \citenamefont {Parshall},
  \citenamefont {Dmowski}, \citenamefont {Lamago}, \citenamefont {Heid},
  \citenamefont {Bohnen}, \citenamefont {Sefat}, \citenamefont {McGuire},
  \citenamefont {Sales}, \citenamefont {Mandrus}, \citenamefont {Subedi},
  \citenamefont {Singh}, \citenamefont {Alatas}, \citenamefont {Upton},
  \citenamefont {Said}, \citenamefont {Cunsolo}, \citenamefont {Shvyd'ko},\
  and\ \citenamefont {Egami}}]{ReznikPRB2009}%
  \BibitemOpen
  \bibfield  {author} {\bibinfo {author} {\bibfnamefont {D.}~\bibnamefont
  {Reznik}}, \bibinfo {author} {\bibfnamefont {K.}~\bibnamefont {Lokshin}},
  \bibinfo {author} {\bibfnamefont {D.~C.}\ \bibnamefont {Mitchell}}, \bibinfo
  {author} {\bibfnamefont {D.}~\bibnamefont {Parshall}}, \bibinfo {author}
  {\bibfnamefont {W.}~\bibnamefont {Dmowski}}, \bibinfo {author} {\bibfnamefont
  {D.}~\bibnamefont {Lamago}}, \bibinfo {author} {\bibfnamefont
  {R.}~\bibnamefont {Heid}}, \bibinfo {author} {\bibfnamefont {K.-P.}\
  \bibnamefont {Bohnen}}, \bibinfo {author} {\bibfnamefont {A.~S.}\
  \bibnamefont {Sefat}}, \bibinfo {author} {\bibfnamefont {M.~A.}\ \bibnamefont
  {McGuire}}, \bibinfo {author} {\bibfnamefont {B.~C.}\ \bibnamefont {Sales}},
  \bibinfo {author} {\bibfnamefont {D.~G.}\ \bibnamefont {Mandrus}}, \bibinfo
  {author} {\bibfnamefont {A.}~\bibnamefont {Subedi}}, \bibinfo {author}
  {\bibfnamefont {D.~J.}\ \bibnamefont {Singh}}, \bibinfo {author}
  {\bibfnamefont {A.}~\bibnamefont {Alatas}}, \bibinfo {author} {\bibfnamefont
  {M.~H.}\ \bibnamefont {Upton}}, \bibinfo {author} {\bibfnamefont {A.~H.}\
  \bibnamefont {Said}}, \bibinfo {author} {\bibfnamefont {A.}~\bibnamefont
  {Cunsolo}}, \bibinfo {author} {\bibfnamefont {Yu.}\ \bibnamefont {Shvyd'ko}},
  \ and\ \bibinfo {author} {\bibfnamefont {T.}~\bibnamefont {Egami}},\
  }\bibfield  {title} {\enquote {\bibinfo {title} {Phonons in doped and undoped
  {B}a{F}e$_2${A}s$_2$ investigated by inelastic x-ray scattering},}\ }\href
  {\doibase 10.1103/PhysRevB.80.214534} {\bibfield  {journal} {\bibinfo
  {journal} {Phys. Rev. B}\ }\textbf {\bibinfo {volume} {80}},\ \bibinfo
  {pages} {214534} (\bibinfo {year} {2009})}\BibitemShut {NoStop}%
\bibitem [{\citenamefont {Hahn}\ \emph {et~al.}(2009)\citenamefont {Hahn},
  \citenamefont {Lee}, \citenamefont {Ni}, \citenamefont {Canfield},
  \citenamefont {Goldman}, \citenamefont {McQueeney}, \citenamefont {Harmon},
  \citenamefont {Alatas}, \citenamefont {Leu}, \citenamefont {Alp},
  \citenamefont {Chung}, \citenamefont {Todorov},\ and\ \citenamefont
  {Kanatzidis}}]{Hahn2009PRB}%
  \BibitemOpen
  \bibfield  {author} {\bibinfo {author} {\bibfnamefont {S.~E.}\ \bibnamefont
  {Hahn}}, \bibinfo {author} {\bibfnamefont {Y.}~\bibnamefont {Lee}}, \bibinfo
  {author} {\bibfnamefont {N.}~\bibnamefont {Ni}}, \bibinfo {author}
  {\bibfnamefont {P.~C.}\ \bibnamefont {Canfield}}, \bibinfo {author}
  {\bibfnamefont {A.~I.}\ \bibnamefont {Goldman}}, \bibinfo {author}
  {\bibfnamefont {R.~J.}\ \bibnamefont {McQueeney}}, \bibinfo {author}
  {\bibfnamefont {B.~N.}\ \bibnamefont {Harmon}}, \bibinfo {author}
  {\bibfnamefont {A.}~\bibnamefont {Alatas}}, \bibinfo {author} {\bibfnamefont
  {B.~M.}\ \bibnamefont {Leu}}, \bibinfo {author} {\bibfnamefont {E.~E.}\
  \bibnamefont {Alp}}, \bibinfo {author} {\bibfnamefont {D.~Y.}\ \bibnamefont
  {Chung}}, \bibinfo {author} {\bibfnamefont {I.~S.}\ \bibnamefont {Todorov}},
  \ and\ \bibinfo {author} {\bibfnamefont {M.~G.}\ \bibnamefont {Kanatzidis}},\
  }\bibfield  {title} {\enquote {\bibinfo {title} {Influence of magnetism on
  phonons in {C}a{F}e$_2${A}s$_2$ as seen via inelastic x-ray scattering},}\
  }\href {\doibase 10.1103/PhysRevB.79.220511} {\bibfield  {journal} {\bibinfo
  {journal} {Phys. Rev. B}\ }\textbf {\bibinfo {volume} {79}},\ \bibinfo
  {pages} {220511} (\bibinfo {year} {2009})}\BibitemShut {NoStop}%
\bibitem [{\citenamefont {Mittal}\ \emph {et~al.}(2013)\citenamefont {Mittal},
  \citenamefont {Gupta}, \citenamefont {Chaplot}, \citenamefont {Zbiri},
  \citenamefont {Rols}, \citenamefont {Schober}, \citenamefont {Su},
  \citenamefont {Brueckel},\ and\ \citenamefont {Wolf}}]{Mittal2013PRB}%
  \BibitemOpen
  \bibfield  {author} {\bibinfo {author} {\bibfnamefont {R.}~\bibnamefont
  {Mittal}}, \bibinfo {author} {\bibfnamefont {M.~K.}\ \bibnamefont {Gupta}},
  \bibinfo {author} {\bibfnamefont {S.~L.}\ \bibnamefont {Chaplot}}, \bibinfo
  {author} {\bibfnamefont {M.}~\bibnamefont {Zbiri}}, \bibinfo {author}
  {\bibfnamefont {S.}~\bibnamefont {Rols}}, \bibinfo {author} {\bibfnamefont
  {H.}~\bibnamefont {Schober}}, \bibinfo {author} {\bibfnamefont
  {Y.}~\bibnamefont {Su}}, \bibinfo {author} {\bibfnamefont {Th.}\ \bibnamefont
  {Brueckel}}, \ and\ \bibinfo {author} {\bibfnamefont {T.}~\bibnamefont
  {Wolf}},\ }\bibfield  {title} {\enquote {\bibinfo {title} {Spin-phonon
  coupling in {K}${}_{0.8}${F}e${}_{1.6}${S}e${}_{2}$ and
  {K}{F}e${}_{2}${S}e${}_{2}$: Inelastic neutron scattering and $ab$ initio
  phonon calculations},}\ }\href {\doibase 10.1103/PhysRevB.87.184502}
  {\bibfield  {journal} {\bibinfo  {journal} {Phys. Rev. B}\ }\textbf {\bibinfo
  {volume} {87}},\ \bibinfo {pages} {184502} (\bibinfo {year}
  {2013})}\BibitemShut {NoStop}%
\bibitem [{\citenamefont {Hahn}\ \emph {et~al.}(2013)\citenamefont {Hahn},
  \citenamefont {Tucker}, \citenamefont {Yan}, \citenamefont {Said},
  \citenamefont {Leu}, \citenamefont {McCallum}, \citenamefont {Alp},
  \citenamefont {Lograsso}, \citenamefont {McQueeney},\ and\ \citenamefont
  {Harmon}}]{Hahn2013PRB}%
  \BibitemOpen
  \bibfield  {author} {\bibinfo {author} {\bibfnamefont {S.~E.}\ \bibnamefont
  {Hahn}}, \bibinfo {author} {\bibfnamefont {G.~S.}\ \bibnamefont {Tucker}},
  \bibinfo {author} {\bibfnamefont {J.-Q.}\ \bibnamefont {Yan}}, \bibinfo
  {author} {\bibfnamefont {A.~H.}\ \bibnamefont {Said}}, \bibinfo {author}
  {\bibfnamefont {B.~M.}\ \bibnamefont {Leu}}, \bibinfo {author} {\bibfnamefont
  {R.~W.}\ \bibnamefont {McCallum}}, \bibinfo {author} {\bibfnamefont {E.~E.}\
  \bibnamefont {Alp}}, \bibinfo {author} {\bibfnamefont {T.~A.}\ \bibnamefont
  {Lograsso}}, \bibinfo {author} {\bibfnamefont {R.~J.}\ \bibnamefont
  {McQueeney}}, \ and\ \bibinfo {author} {\bibfnamefont {B.~N.}\ \bibnamefont
  {Harmon}},\ }\bibfield  {title} {\enquote {\bibinfo {title}
  {Magnetism-dependent phonon anomaly in {L}a{F}e{A}s{O} observed via inelastic
  x-ray scattering},}\ }\href {\doibase 10.1103/PhysRevB.87.104518} {\bibfield
  {journal} {\bibinfo  {journal} {Phys. Rev. B}\ }\textbf {\bibinfo {volume}
  {87}},\ \bibinfo {pages} {104518} (\bibinfo {year} {2013})}\BibitemShut
  {NoStop}%
\bibitem [{\citenamefont {Choi}\ \emph {et~al.}(2008)\citenamefont {Choi},
  \citenamefont {Wulferding}, \citenamefont {Lemmens}, \citenamefont {Ni},
  \citenamefont {Bud'ko},\ and\ \citenamefont {Canfield}}]{Choi2008PhysRevB78}%
  \BibitemOpen
  \bibfield  {author} {\bibinfo {author} {\bibfnamefont {K.~Y.}\ \bibnamefont
  {Choi}}, \bibinfo {author} {\bibfnamefont {D.}~\bibnamefont {Wulferding}},
  \bibinfo {author} {\bibfnamefont {P.}~\bibnamefont {Lemmens}}, \bibinfo
  {author} {\bibfnamefont {N.}~\bibnamefont {Ni}}, \bibinfo {author}
  {\bibfnamefont {S.~L.}\ \bibnamefont {Bud'ko}}, \ and\ \bibinfo {author}
  {\bibfnamefont {P.~C.}\ \bibnamefont {Canfield}},\ }\bibfield  {title}
  {\enquote {\bibinfo {title} {Lattice and electronic anomalies of
  {C}a{F}e$_{2}${A}s$_{2}$ studied by {R}aman spectroscopy},}\ }\href {\doibase
  10.1103/PhysRevB.78.212503} {\bibfield  {journal} {\bibinfo  {journal} {Phys.
  Rev. B}\ }\textbf {\bibinfo {volume} {78}},\ \bibinfo {pages} {212503}
  (\bibinfo {year} {2008})}\BibitemShut {NoStop}%
\bibitem [{\citenamefont {Zhang}\ \emph
  {et~al.}(2014{\natexlab{b}})\citenamefont {Zhang}, \citenamefont {Richard},
  \citenamefont {Ding}, \citenamefont {Sefat}, \citenamefont {Gillett},
  \citenamefont {Sebastian}, \citenamefont {Khodas},\ and\ \citenamefont
  {Blumberg}}]{ZhangWL2014arxiv}%
  \BibitemOpen
  \bibfield  {author} {\bibinfo {author} {\bibfnamefont {W.~L.}\ \bibnamefont
  {Zhang}}, \bibinfo {author} {\bibfnamefont {P.}~\bibnamefont {Richard}},
  \bibinfo {author} {\bibfnamefont {H.}~\bibnamefont {Ding}}, \bibinfo {author}
  {\bibfnamefont {A.~S.}\ \bibnamefont {Sefat}}, \bibinfo {author}
  {\bibfnamefont {J.}~\bibnamefont {Gillett}}, \bibinfo {author} {\bibfnamefont
  {S.~E.}\ \bibnamefont {Sebastian}}, \bibinfo {author} {\bibfnamefont
  {M.}~\bibnamefont {Khodas}}, \ and\ \bibinfo {author} {\bibfnamefont
  {G.}~\bibnamefont {Blumberg}},\ }\bibfield  {title} {\enquote {\bibinfo
  {title} {On the origin of the electronic anisotropy in iron pnicitde
  superconductors},}\ }\href {https://arxiv.org/abs/1410.6452} {\bibfield
  {journal} {\bibinfo  {journal} {arXiv:1410.6452}\ } (\bibinfo {year}
  {2014}{\natexlab{b}})}\BibitemShut {NoStop}%
\bibitem [{\citenamefont {Chauvi\`ere}\ \emph {et~al.}(2011)\citenamefont
  {Chauvi\`ere}, \citenamefont {Gallais}, \citenamefont {Cazayous},
  \citenamefont {M\'easson}, \citenamefont {Sacuto}, \citenamefont {Colson},\
  and\ \citenamefont {Forget}}]{Chauvilere_PRB80}%
  \BibitemOpen
  \bibfield  {author} {\bibinfo {author} {\bibfnamefont {L.}~\bibnamefont
  {Chauvi\`ere}}, \bibinfo {author} {\bibfnamefont {Y.}~\bibnamefont
  {Gallais}}, \bibinfo {author} {\bibfnamefont {M.}~\bibnamefont {Cazayous}},
  \bibinfo {author} {\bibfnamefont {M.~A.}\ \bibnamefont {M\'easson}}, \bibinfo
  {author} {\bibfnamefont {A.}~\bibnamefont {Sacuto}}, \bibinfo {author}
  {\bibfnamefont {D.}~\bibnamefont {Colson}}, \ and\ \bibinfo {author}
  {\bibfnamefont {A.}~\bibnamefont {Forget}},\ }\bibfield  {title} {\enquote
  {\bibinfo {title} {Raman scattering study of spin-density-wave order and
  electron-phonon coupling in {B}a({F}e$_{1-x}${C}o$_{x}$)$_{2}${A}s$_{2}$},}\
  }\href {\doibase 10.1103/PhysRevB.84.104508} {\bibfield  {journal} {\bibinfo
  {journal} {Phys. Rev. B}\ }\textbf {\bibinfo {volume} {84}},\ \bibinfo
  {pages} {104508} (\bibinfo {year} {2011})}\BibitemShut {NoStop}%
\bibitem [{\citenamefont {Kretzschmar}\ \emph {et~al.}(2016)\citenamefont
  {Kretzschmar}, \citenamefont {Bohm}, \citenamefont {Karahasanovic},
  \citenamefont {Muschler}, \citenamefont {Baum}, \citenamefont {Jost},
  \citenamefont {Schmalian}, \citenamefont {Caprara}, \citenamefont {Grilli},
  \citenamefont {Di~Castro}, \citenamefont {Analytis}, \citenamefont {Chu},
  \citenamefont {Fisher},\ and\ \citenamefont {Hackl}}]{Kretzschmar2016NatPhy}%
  \BibitemOpen
  \bibfield  {author} {\bibinfo {author} {\bibfnamefont {F.}~\bibnamefont
  {Kretzschmar}}, \bibinfo {author} {\bibfnamefont {T.}~\bibnamefont {Bohm}},
  \bibinfo {author} {\bibfnamefont {U.}~\bibnamefont {Karahasanovic}}, \bibinfo
  {author} {\bibfnamefont {B.}~\bibnamefont {Muschler}}, \bibinfo {author}
  {\bibfnamefont {A.}~\bibnamefont {Baum}}, \bibinfo {author} {\bibfnamefont
  {D.}~\bibnamefont {Jost}}, \bibinfo {author} {\bibfnamefont {J.}~\bibnamefont
  {Schmalian}}, \bibinfo {author} {\bibfnamefont {S.}~\bibnamefont {Caprara}},
  \bibinfo {author} {\bibfnamefont {M.}~\bibnamefont {Grilli}}, \bibinfo
  {author} {\bibfnamefont {C.}~\bibnamefont {Di~Castro}}, \bibinfo {author}
  {\bibfnamefont {J.~G.}\ \bibnamefont {Analytis}}, \bibinfo {author}
  {\bibfnamefont {J.-H.}\ \bibnamefont {Chu}}, \bibinfo {author} {\bibfnamefont
  {I.~R.}\ \bibnamefont {Fisher}}, \ and\ \bibinfo {author} {\bibfnamefont
  {R.}~\bibnamefont {Hackl}},\ }\bibfield  {title} {\enquote {\bibinfo {title}
  {Critical spin fluctuations and the origin of nematic order in
  {B}a({F}e$_{1-x}${C}o$_{x}$)$_2${A}s$_2$},}\ }\href
  {http://dx.doi.org/10.1038/nphys3634} {\bibfield  {journal} {\bibinfo
  {journal} {Nat. Phys.}\ }\textbf {\bibinfo {volume} {12}},\ \bibinfo {pages}
  {560} (\bibinfo {year} {2016})}\BibitemShut {NoStop}%
\bibitem [{\citenamefont {Sugai}\ \emph {et~al.}(2012)\citenamefont {Sugai},
  \citenamefont {Mizuno}, \citenamefont {Watanabe}, \citenamefont {Kawaguchi},
  \citenamefont {Takenaka}, \citenamefont {Ikuta}, \citenamefont {Takayanagi},
  \citenamefont {Hayamizu},\ and\ \citenamefont {Sone}}]{Sugai2012JPSJ}%
  \BibitemOpen
  \bibfield  {author} {\bibinfo {author} {\bibfnamefont {S.}~\bibnamefont
  {Sugai}}, \bibinfo {author} {\bibfnamefont {Y.}~\bibnamefont {Mizuno}},
  \bibinfo {author} {\bibfnamefont {R.}~\bibnamefont {Watanabe}}, \bibinfo
  {author} {\bibfnamefont {T.}~\bibnamefont {Kawaguchi}}, \bibinfo {author}
  {\bibfnamefont {K.}~\bibnamefont {Takenaka}}, \bibinfo {author}
  {\bibfnamefont {H.}~\bibnamefont {Ikuta}}, \bibinfo {author} {\bibfnamefont
  {Y.}~\bibnamefont {Takayanagi}}, \bibinfo {author} {\bibfnamefont
  {N.}~\bibnamefont {Hayamizu}}, \ and\ \bibinfo {author} {\bibfnamefont
  {Y.}~\bibnamefont {Sone}},\ }\bibfield  {title} {\enquote {\bibinfo {title}
  {Spin-density-wave gap with {D}irac nodes and two-magnon raman scattering in
  {B}a{F}e$_2${A}s$_2$},}\ }\href {\doibase 10.1143/JPSJ.81.024718} {\bibfield
  {journal} {\bibinfo  {journal} {J. Phys. Soc. Jpn.}\ }\textbf {\bibinfo
  {volume} {81}},\ \bibinfo {pages} {024718} (\bibinfo {year}
  {2012})}\BibitemShut {NoStop}%
\bibitem [{\citenamefont {Wu}\ \emph {et~al.}()\citenamefont {Wu},
  \citenamefont {Zhang}, \citenamefont {Li}, \citenamefont {Cao}, \citenamefont
  {Kung}, \citenamefont {Ding}, \citenamefont {Richard},\ and\ \citenamefont
  {Blumberg}}]{Au_paper}%
  \BibitemOpen
  \bibfield  {author} {\bibinfo {author} {\bibfnamefont {S.-F.}\ \bibnamefont
  {Wu}}, \bibinfo {author} {\bibfnamefont {W.-L.}\ \bibnamefont {Zhang}},
  \bibinfo {author} {\bibfnamefont {L.}~\bibnamefont {Li}}, \bibinfo {author}
  {\bibfnamefont {H.~B.}\ \bibnamefont {Cao}}, \bibinfo {author} {\bibfnamefont
  {Sefat A.~S.}\ \bibnamefont {Kung}, \bibfnamefont {H.-H.}}, \bibinfo {author}
  {\bibfnamefont {H.}~\bibnamefont {Ding}}, \bibinfo {author} {\bibfnamefont
  {P.}~\bibnamefont {Richard}}, \ and\ \bibinfo {author} {\bibfnamefont
  {G.}~\bibnamefont {Blumberg}},\ }\bibfield  {title} {\enquote {\bibinfo
  {title} {{Anomalous magneto-elastic coupling in Au-doped BaFe$_2$As$_2$}},}\
  }\href@noop {} {\bibinfo  {journal} {Submitted to PRB}\ }\BibitemShut {NoStop}%
\bibitem [{\citenamefont {Kaneko}\ \emph {et~al.}(2008)\citenamefont {Kaneko},
  \citenamefont {Gomes}, \citenamefont {Garcia-Flores}, \citenamefont {Yan},
  \citenamefont {Lograsso}, \citenamefont {Barberis}, \citenamefont {Vaknin},\
  and\ \citenamefont {Granado}}]{Kaneko2017arxiv}%
  \BibitemOpen
\bibfield  {journal} {  }\bibfield  {author} {\bibinfo {author} {\bibfnamefont
  {U.~F.}\ \bibnamefont {Kaneko}}, \bibinfo {author} {\bibfnamefont {P.~F.}\
  \bibnamefont {Gomes}}, \bibinfo {author} {\bibfnamefont {A.~F.}\ \bibnamefont
  {Garcia-Flores}}, \bibinfo {author} {\bibfnamefont {J.~Q.}\ \bibnamefont
  {Yan}}, \bibinfo {author} {\bibfnamefont {T.~A.}\ \bibnamefont {Lograsso}},
  \bibinfo {author} {\bibfnamefont {G.~E.}\ \bibnamefont {Barberis}}, \bibinfo
  {author} {\bibfnamefont {D.}~\bibnamefont {Vaknin}}, \ and\ \bibinfo {author}
  {\bibfnamefont {E.}~\bibnamefont {Granado}},\ }\bibfield  {title} {\enquote
  {\bibinfo {title} {Nematic fluctuations and phase transitions in
  {L}a{F}e{A}s{O}: a {R}aman scattering study},}\ }\href {\doibase
  10.1103/PhysRevLett.101.257005} {\bibfield  {journal} {\bibinfo  {journal}
  {Phys. Rev. B}\ }\textbf {\bibinfo {volume} {96}},\ \bibinfo {pages} {257005}
  (\bibinfo {year} {2008})}\BibitemShut {NoStop}%
\bibitem [{\citenamefont {Egami}\ \emph {et~al.}(2010)\citenamefont {Egami},
  \citenamefont {Fine}, \citenamefont {Parshall}, \citenamefont {Subedi},\ and\
  \citenamefont {Singh}}]{Egami2010JAdvances}%
  \BibitemOpen
  \bibfield  {author} {\bibinfo {author} {\bibfnamefont {T.}~\bibnamefont
  {Egami}}, \bibinfo {author} {\bibfnamefont {B.~V.}\ \bibnamefont {Fine}},
  \bibinfo {author} {\bibfnamefont {D.}~\bibnamefont {Parshall}}, \bibinfo
  {author} {\bibfnamefont {A.}~\bibnamefont {Subedi}}, \ and\ \bibinfo {author}
  {\bibfnamefont {D.~J.}\ \bibnamefont {Singh}},\ }\bibfield  {title} {\enquote
  {\bibinfo {title} {Spin-lattice coupling and superconductivity in {F}e
  pnictides},}\ }\href {\doibase 10.1155/2010/164916} {\bibfield  {journal}
  {\bibinfo  {journal} {J. Adv. Cond. Matter Phys.}\ }\textbf {\bibinfo
  {volume} {7}},\ \bibinfo {pages} {164916} (\bibinfo {year}
  {2010})}\BibitemShut {NoStop}%
\bibitem [{\citenamefont {Gerber}\ \emph {et~al.}(2017)\citenamefont {Gerber},
  \citenamefont {Yang}, \citenamefont {Zhu}, \citenamefont {Soifer},
  \citenamefont {Sobota}, \citenamefont {Rebec}, \citenamefont {Lee},
  \citenamefont {Jia}, \citenamefont {Moritz}, \citenamefont {Jia},
  \citenamefont {Gauthier}, \citenamefont {Li}, \citenamefont {Leuenberger},
  \citenamefont {Zhang}, \citenamefont {Chaix}, \citenamefont {Li},
  \citenamefont {Jang}, \citenamefont {Lee}, \citenamefont {Yi}, \citenamefont
  {Dakovski}, \citenamefont {Song}, \citenamefont {Glownia}, \citenamefont
  {Nelson}, \citenamefont {Kim}, \citenamefont {Chuang}, \citenamefont
  {Hussain}, \citenamefont {Moore}, \citenamefont {Devereaux}, \citenamefont
  {Lee}, \citenamefont {Kirchmann},\ and\ \citenamefont
  {Shen}}]{Gerber2017Science}%
  \BibitemOpen
  \bibfield  {author} {\bibinfo {author} {\bibfnamefont {S.}~\bibnamefont
  {Gerber}}, \bibinfo {author} {\bibfnamefont {S.-L.}\ \bibnamefont {Yang}},
  \bibinfo {author} {\bibfnamefont {D.}~\bibnamefont {Zhu}}, \bibinfo {author}
  {\bibfnamefont {H.}~\bibnamefont {Soifer}}, \bibinfo {author} {\bibfnamefont
  {J.~A.}\ \bibnamefont {Sobota}}, \bibinfo {author} {\bibfnamefont
  {S.}~\bibnamefont {Rebec}}, \bibinfo {author} {\bibfnamefont {J.~J.}\
  \bibnamefont {Lee}}, \bibinfo {author} {\bibfnamefont {T.}~\bibnamefont
  {Jia}}, \bibinfo {author} {\bibfnamefont {B.}~\bibnamefont {Moritz}},
  \bibinfo {author} {\bibfnamefont {C.}~\bibnamefont {Jia}}, \bibinfo {author}
  {\bibfnamefont {A.}~\bibnamefont {Gauthier}}, \bibinfo {author}
  {\bibfnamefont {Y.}~\bibnamefont {Li}}, \bibinfo {author} {\bibfnamefont
  {D.}~\bibnamefont {Leuenberger}}, \bibinfo {author} {\bibfnamefont
  {Y.}~\bibnamefont {Zhang}}, \bibinfo {author} {\bibfnamefont
  {L.}~\bibnamefont {Chaix}}, \bibinfo {author} {\bibfnamefont
  {W.}~\bibnamefont {Li}}, \bibinfo {author} {\bibfnamefont {H.}~\bibnamefont
  {Jang}}, \bibinfo {author} {\bibfnamefont {J.-S.}\ \bibnamefont {Lee}},
  \bibinfo {author} {\bibfnamefont {M.}~\bibnamefont {Yi}}, \bibinfo {author}
  {\bibfnamefont {G.~L.}\ \bibnamefont {Dakovski}}, \bibinfo {author}
  {\bibfnamefont {S.}~\bibnamefont {Song}}, \bibinfo {author} {\bibfnamefont
  {J.~M.}\ \bibnamefont {Glownia}}, \bibinfo {author} {\bibfnamefont
  {S.}~\bibnamefont {Nelson}}, \bibinfo {author} {\bibfnamefont {K.~W.}\
  \bibnamefont {Kim}}, \bibinfo {author} {\bibfnamefont {Y.-D.}\ \bibnamefont
  {Chuang}}, \bibinfo {author} {\bibfnamefont {Z.}~\bibnamefont {Hussain}},
  \bibinfo {author} {\bibfnamefont {R.~G.}\ \bibnamefont {Moore}}, \bibinfo
  {author} {\bibfnamefont {T.~P.}\ \bibnamefont {Devereaux}}, \bibinfo {author}
  {\bibfnamefont {W.-S.}\ \bibnamefont {Lee}}, \bibinfo {author} {\bibfnamefont
  {P.~S.}\ \bibnamefont {Kirchmann}}, \ and\ \bibinfo {author} {\bibfnamefont
  {Z.-X.}\ \bibnamefont {Shen}},\ }\bibfield  {title} {\enquote {\bibinfo
  {title} {Femtosecond electron-phonon lock-in by photoemission and x-ray
  free-electron laser},}\ }\href {\doibase 10.1126/science.aak9946} {\bibfield
  {journal} {\bibinfo  {journal} {Science}\ }\textbf {\bibinfo {volume}
  {357}},\ \bibinfo {pages} {71} (\bibinfo {year} {2017})}\BibitemShut
  {NoStop}%
\bibitem [{\citenamefont {Sefat}(2013)}]{Sefat2013bulk}%
  \BibitemOpen
  \bibfield  {author} {\bibinfo {author} {\bibfnamefont {A.~S.}\ \bibnamefont
  {Sefat}},\ }\bibfield  {title} {\enquote {\bibinfo {title} {Bulk synthesis of
  iron-based superconductors},}\ }\href
  {http://www.sciencedirect.com/science/article/pii/S1359028613000314}
  {\bibfield  {journal} {\bibinfo  {journal} {Curr. Opin. Solid State Mater.
  Sci.}\ }\textbf {\bibinfo {volume} {17}},\ \bibinfo {pages} {59--64}
  (\bibinfo {year} {2013})}\BibitemShut {NoStop}%
\bibitem [{\citenamefont {Li}\ \emph {et~al.}(2015)\citenamefont {Li},
  \citenamefont {Cao}, \citenamefont {McGuire}, \citenamefont {Kim},
  \citenamefont {Stewart},\ and\ \citenamefont {Sefat}}]{Li2015PRB}%
  \BibitemOpen
  \bibfield  {author} {\bibinfo {author} {\bibfnamefont {L.}~\bibnamefont
  {Li}}, \bibinfo {author} {\bibfnamefont {H.~B.}\ \bibnamefont {Cao}},
  \bibinfo {author} {\bibfnamefont {M.~A.}\ \bibnamefont {McGuire}}, \bibinfo
  {author} {\bibfnamefont {J.~S.}\ \bibnamefont {Kim}}, \bibinfo {author}
  {\bibfnamefont {G.~R.}\ \bibnamefont {Stewart}}, \ and\ \bibinfo {author}
  {\bibfnamefont {A.~S.}\ \bibnamefont {Sefat}},\ }\bibfield  {title} {\enquote
  {\bibinfo {title} {Role of magnetism in superconductivity of
  {B}a{F}e$_2${A}s$_2$: Study of $5d$ {A}u-doped crystals},}\ }\href {\doibase
  10.1103/PhysRevB.92.094504} {\bibfield  {journal} {\bibinfo  {journal} {Phys.
  Rev. B}\ }\textbf {\bibinfo {volume} {92}},\ \bibinfo {pages} {094504}
  (\bibinfo {year} {2015})}\BibitemShut {NoStop}%
\bibitem [{\citenamefont {Tanatar}\ \emph {et~al.}(2012)\citenamefont
  {Tanatar}, \citenamefont {Spyrison}, \citenamefont {Cho}, \citenamefont
  {Blomberg}, \citenamefont {Tan}, \citenamefont {Dai}, \citenamefont {Zhang},\
  and\ \citenamefont {Prozorov}}]{Tanatar2012PRB}%
  \BibitemOpen
  \bibfield  {author} {\bibinfo {author} {\bibfnamefont {M.~A.}\ \bibnamefont
  {Tanatar}}, \bibinfo {author} {\bibfnamefont {N.}~\bibnamefont {Spyrison}},
  \bibinfo {author} {\bibfnamefont {Kyuil}\ \bibnamefont {Cho}}, \bibinfo
  {author} {\bibfnamefont {E.~C.}\ \bibnamefont {Blomberg}}, \bibinfo {author}
  {\bibfnamefont {Guotai}\ \bibnamefont {Tan}}, \bibinfo {author}
  {\bibfnamefont {Pengcheng}\ \bibnamefont {Dai}}, \bibinfo {author}
  {\bibfnamefont {Chenglin}\ \bibnamefont {Zhang}}, \ and\ \bibinfo {author}
  {\bibfnamefont {R.}~\bibnamefont {Prozorov}},\ }\bibfield  {title} {\enquote
  {\bibinfo {title} {Evolution of normal and superconducting properties of
  single crystals of {N}a${}_{1-\ensuremath{\delta}}${F}e{A}s upon interaction
  with environment},}\ }\href {\doibase 10.1103/PhysRevB.85.014510} {\bibfield
  {journal} {\bibinfo  {journal} {Phys. Rev. B}\ }\textbf {\bibinfo {volume}
  {85}},\ \bibinfo {pages} {014510} (\bibinfo {year} {2012})}\BibitemShut
  {NoStop}%
\bibitem [{\citenamefont {Kamihara}\ \emph {et~al.}(2008)\citenamefont
  {Kamihara}, \citenamefont {Watanabe}, \citenamefont {Hirano},\ and\
  \citenamefont {Hosono}}]{Kamihara2008JACS}%
  \BibitemOpen
  \bibfield  {author} {\bibinfo {author} {\bibfnamefont {Y.}~\bibnamefont
  {Kamihara}}, \bibinfo {author} {\bibfnamefont {T.}~\bibnamefont {Watanabe}},
  \bibinfo {author} {\bibfnamefont {M.}~\bibnamefont {Hirano}}, \ and\ \bibinfo
  {author} {\bibfnamefont {H.}~\bibnamefont {Hosono}},\ }\bibfield  {title}
  {\enquote {\bibinfo {title} {Iron-based layered superconductor
  {L}a{O}$_{1-x}${F}$_x${F}e{A}s (x = $0.05-0.12$) with ${T}_c$ = 26 {K}},}\
  }\href {\doibase 10.1021/ja800073m} {\bibfield  {journal} {\bibinfo
  {journal} {J. Am. Chem. Soc}\ }\textbf {\bibinfo {volume} {130}},\ \bibinfo
  {pages} {3296} (\bibinfo {year} {2008})}\BibitemShut {NoStop}%
\bibitem [{\citenamefont {de~la Cruz}\ \emph
  {et~al.}(2008{\natexlab{a}})\citenamefont {de~la Cruz}, \citenamefont
  {Huang}, \citenamefont {Lynn}, \citenamefont {Li}, \citenamefont {II},
  \citenamefont {Zarestky}, \citenamefont {Mook}, \citenamefont {Chen},
  \citenamefont {Luo}, \citenamefont {Wang},\ and\ \citenamefont
  {Dai}}]{Clarina2008Nature}%
  \BibitemOpen
  \bibfield  {author} {\bibinfo {author} {\bibfnamefont {C.}~\bibnamefont
  {de~la Cruz}}, \bibinfo {author} {\bibfnamefont {Q.}~\bibnamefont {Huang}},
  \bibinfo {author} {\bibfnamefont {J.~W.}\ \bibnamefont {Lynn}}, \bibinfo
  {author} {\bibfnamefont {Jiying}\ \bibnamefont {Li}}, \bibinfo {author}
  {\bibfnamefont {W.~Ratcliff}\ \bibnamefont {II}}, \bibinfo {author}
  {\bibfnamefont {J.~L.}\ \bibnamefont {Zarestky}}, \bibinfo {author}
  {\bibfnamefont {H.~A.}\ \bibnamefont {Mook}}, \bibinfo {author}
  {\bibfnamefont {G.~F.}\ \bibnamefont {Chen}}, \bibinfo {author}
  {\bibfnamefont {J.~L.}\ \bibnamefont {Luo}}, \bibinfo {author} {\bibfnamefont
  {N.~L.}\ \bibnamefont {Wang}}, \ and\ \bibinfo {author} {\bibfnamefont
  {P.~C}\ \bibnamefont {Dai}},\ }\bibfield  {title} {\enquote {\bibinfo {title}
  {Magnetic order close to superconductivity in the iron-based layered
  {L}a{O}$_{1-x}${F}$_x${F}e{A}s systems},}\ }\href {\doibase
  10.1038/nature07057} {\bibfield  {journal} {\bibinfo  {journal} {Nature}\
  }\textbf {\bibinfo {volume} {453}},\ \bibinfo {pages} {899} (\bibinfo {year}
  {2008}{\natexlab{a}})}\BibitemShut {NoStop}%
\bibitem [{\citenamefont {Hosoi}\ \emph {et~al.}(2016)\citenamefont {Hosoi},
  \citenamefont {Matsuura}, \citenamefont {Ishida}, \citenamefont {Wang},
  \citenamefont {Mizukami}, \citenamefont {Watashige}, \citenamefont
  {Kasahara}, \citenamefont {Matsuda},\ and\ \citenamefont
  {Shibauchi}}]{Hosoi2016PNAS}%
  \BibitemOpen
  \bibfield  {author} {\bibinfo {author} {\bibfnamefont {S.}~\bibnamefont
  {Hosoi}}, \bibinfo {author} {\bibfnamefont {K.}~\bibnamefont {Matsuura}},
  \bibinfo {author} {\bibfnamefont {K.}~\bibnamefont {Ishida}}, \bibinfo
  {author} {\bibfnamefont {H.}~\bibnamefont {Wang}}, \bibinfo {author}
  {\bibfnamefont {Y.}~\bibnamefont {Mizukami}}, \bibinfo {author}
  {\bibfnamefont {T.}~\bibnamefont {Watashige}}, \bibinfo {author}
  {\bibfnamefont {S.}~\bibnamefont {Kasahara}}, \bibinfo {author}
  {\bibfnamefont {Y.}~\bibnamefont {Matsuda}}, \ and\ \bibinfo {author}
  {\bibfnamefont {T.}~\bibnamefont {Shibauchi}},\ }\bibfield  {title} {\enquote
  {\bibinfo {title} {Nematic quantum critical point without magnetism in
  {F}e{S}e$_{1-x}${S}$_x$ superconductors},}\ }\href {\doibase
  10.1073/pnas.1605806113} {\bibfield  {journal} {\bibinfo  {journal} {Proc.
  Natl. Acad. Sci. U.S.A.}\ }\textbf {\bibinfo {volume} {113}},\ \bibinfo
  {pages} {8139} (\bibinfo {year} {2016})}\BibitemShut {NoStop}%
\bibitem [{\citenamefont {Thorsm\o{}lle}\ \emph {et~al.}(2016)\citenamefont
  {Thorsm\o{}lle}, \citenamefont {Khodas}, \citenamefont {Yin}, \citenamefont
  {Zhang}, \citenamefont {Carr}, \citenamefont {Dai},\ and\ \citenamefont
  {Blumberg}}]{Thorsmolle2016PRB}%
  \BibitemOpen
  \bibfield  {author} {\bibinfo {author} {\bibfnamefont {V.~K.}\ \bibnamefont
  {Thorsm\o{}lle}}, \bibinfo {author} {\bibfnamefont {M.}~\bibnamefont
  {Khodas}}, \bibinfo {author} {\bibfnamefont {Z.~P.}\ \bibnamefont {Yin}},
  \bibinfo {author} {\bibfnamefont {Chenglin}\ \bibnamefont {Zhang}}, \bibinfo
  {author} {\bibfnamefont {S.~V.}\ \bibnamefont {Carr}}, \bibinfo {author}
  {\bibfnamefont {Pengcheng}\ \bibnamefont {Dai}}, \ and\ \bibinfo {author}
  {\bibfnamefont {G.}~\bibnamefont {Blumberg}},\ }\bibfield  {title} {\enquote
  {\bibinfo {title} {Critical quadrupole fluctuations and collective modes in
  iron pnictide superconductors},}\ }\href {\doibase
  10.1103/PhysRevB.93.054515} {\bibfield  {journal} {\bibinfo  {journal} {Phys.
  Rev. B}\ }\textbf {\bibinfo {volume} {93}},\ \bibinfo {pages} {054515}
  (\bibinfo {year} {2016})}\BibitemShut {NoStop}%
\bibitem [{\citenamefont {Tegel}\ \emph {et~al.}(2008)\citenamefont {Tegel},
  \citenamefont {Rotter}, \citenamefont {Weiß}, \citenamefont {Schappacher},
  \citenamefont {Pöttgen},\ and\ \citenamefont {Johrendt}}]{Marcus2008JPCM}%
  \BibitemOpen
  \bibfield  {author} {\bibinfo {author} {\bibfnamefont {M.}~\bibnamefont
  {Tegel}}, \bibinfo {author} {\bibfnamefont {M.}~\bibnamefont {Rotter}},
  \bibinfo {author} {\bibfnamefont {V.}~\bibnamefont {Weiß}}, \bibinfo
  {author} {\bibfnamefont {F.~M.}\ \bibnamefont {Schappacher}}, \bibinfo
  {author} {\bibfnamefont {R.}~\bibnamefont {Pöttgen}}, \ and\ \bibinfo
  {author} {\bibfnamefont {D.}~\bibnamefont {Johrendt}},\ }\bibfield  {title}
  {\enquote {\bibinfo {title} {Structural and magnetic phase transitions in the
  ternary iron arsenides {S}r{F}e$_2${A}s$_2$ and {E}u{F}e$_2${A}s$_2$},}\
  }\href {http://stacks.iop.org/0953-8984/20/i=45/a=452201} {\bibfield
  {journal} {\bibinfo  {journal} {J. Phys.: Condens. Matter}\ }\textbf
  {\bibinfo {volume} {20}},\ \bibinfo {pages} {452201} (\bibinfo {year}
  {2008})}\BibitemShut {NoStop}%
\bibitem [{\citenamefont {Xiao}\ \emph {et~al.}(2009)\citenamefont {Xiao},
  \citenamefont {Su}, \citenamefont {Meven}, \citenamefont {Mittal},
  \citenamefont {Kumar}, \citenamefont {Chatterji}, \citenamefont {Price},
  \citenamefont {Persson}, \citenamefont {Kumar}, \citenamefont {Dhar},
  \citenamefont {Thamizhavel},\ and\ \citenamefont
  {Brueckel}}]{Xiao2009EuFe2As2}%
  \BibitemOpen
  \bibfield  {author} {\bibinfo {author} {\bibfnamefont {Y.}~\bibnamefont
  {Xiao}}, \bibinfo {author} {\bibfnamefont {Y.}~\bibnamefont {Su}}, \bibinfo
  {author} {\bibfnamefont {M.}~\bibnamefont {Meven}}, \bibinfo {author}
  {\bibfnamefont {R.}~\bibnamefont {Mittal}}, \bibinfo {author} {\bibfnamefont
  {C.~M.~N.}\ \bibnamefont {Kumar}}, \bibinfo {author} {\bibfnamefont
  {T.}~\bibnamefont {Chatterji}}, \bibinfo {author} {\bibfnamefont
  {S.}~\bibnamefont {Price}}, \bibinfo {author} {\bibfnamefont
  {J.}~\bibnamefont {Persson}}, \bibinfo {author} {\bibfnamefont
  {N.}~\bibnamefont {Kumar}}, \bibinfo {author} {\bibfnamefont {S.~K.}\
  \bibnamefont {Dhar}}, \bibinfo {author} {\bibfnamefont {A.}~\bibnamefont
  {Thamizhavel}}, \ and\ \bibinfo {author} {\bibfnamefont {Th.}\ \bibnamefont
  {Brueckel}},\ }\bibfield  {title} {\enquote {\bibinfo {title} {Magnetic
  structure of {E}u{F}e$_2${A}s$_2$ determined by single-crystal neutron
  diffraction},}\ }\href {\doibase 10.1103/PhysRevB.80.174424} {\bibfield
  {journal} {\bibinfo  {journal} {Phys. Rev. B}\ }\textbf {\bibinfo {volume}
  {80}},\ \bibinfo {pages} {174424} (\bibinfo {year} {2009})}\BibitemShut
  {NoStop}%
\bibitem [{\citenamefont {Huang}\ \emph {et~al.}(2008)\citenamefont {Huang},
  \citenamefont {Qiu}, \citenamefont {Bao}, \citenamefont {Green},
  \citenamefont {Lynn}, \citenamefont {Gasparovic}, \citenamefont {Wu},
  \citenamefont {Wu},\ and\ \citenamefont {Chen}}]{Huang2008PRL}%
  \BibitemOpen
  \bibfield  {author} {\bibinfo {author} {\bibfnamefont {Q.}~\bibnamefont
  {Huang}}, \bibinfo {author} {\bibfnamefont {Y.}~\bibnamefont {Qiu}}, \bibinfo
  {author} {\bibfnamefont {Wei}\ \bibnamefont {Bao}}, \bibinfo {author}
  {\bibfnamefont {M.~A.}\ \bibnamefont {Green}}, \bibinfo {author}
  {\bibfnamefont {J.~W.}\ \bibnamefont {Lynn}}, \bibinfo {author}
  {\bibfnamefont {Y.~C.}\ \bibnamefont {Gasparovic}}, \bibinfo {author}
  {\bibfnamefont {T.}~\bibnamefont {Wu}}, \bibinfo {author} {\bibfnamefont
  {G.}~\bibnamefont {Wu}}, \ and\ \bibinfo {author} {\bibfnamefont {X.~H.}\
  \bibnamefont {Chen}},\ }\bibfield  {title} {\enquote {\bibinfo {title}
  {Neutron-diffraction measurements of magnetic order and a structural
  transition in the parent {B}a{F}e$_{2}${A}s$_{2}$ compound of {F}e{A}s-based
  high-temperature superconductors},}\ }\href {\doibase
  10.1103/PhysRevLett.101.257003} {\bibfield  {journal} {\bibinfo  {journal}
  {Phys. Rev. Lett.}\ }\textbf {\bibinfo {volume} {101}},\ \bibinfo {pages}
  {257003} (\bibinfo {year} {2008})}\BibitemShut {NoStop}%
\bibitem [{\citenamefont {Dai}(2015)}]{Dai2015RevModPhys}%
  \BibitemOpen
  \bibfield  {author} {\bibinfo {author} {\bibfnamefont {P.~C.}\ \bibnamefont
  {Dai}},\ }\bibfield  {title} {\enquote {\bibinfo {title} {Antiferromagnetic
  order and spin dynamics in iron-based superconductors},}\ }\href {\doibase
  10.1103/RevModPhys.87.855} {\bibfield  {journal} {\bibinfo  {journal} {Rev.
  Mod. Phys.}\ }\textbf {\bibinfo {volume} {87}},\ \bibinfo {pages} {855}
  (\bibinfo {year} {2015})}\BibitemShut {NoStop}%
\bibitem [{\citenamefont {Li}\ \emph {et~al.}(2009)\citenamefont {Li},
  \citenamefont {de~la Cruz}, \citenamefont {Huang}, \citenamefont {Chen},
  \citenamefont {Xia}, \citenamefont {Luo}, \citenamefont {Wang},\ and\
  \citenamefont {Dai}}]{Li2009PRB}%
  \BibitemOpen
  \bibfield  {author} {\bibinfo {author} {\bibfnamefont {S.~L}\ \bibnamefont
  {Li}}, \bibinfo {author} {\bibfnamefont {C.}~\bibnamefont {de~la Cruz}},
  \bibinfo {author} {\bibfnamefont {Q.}~\bibnamefont {Huang}}, \bibinfo
  {author} {\bibfnamefont {G.~F.}\ \bibnamefont {Chen}}, \bibinfo {author}
  {\bibfnamefont {T.-L.}\ \bibnamefont {Xia}}, \bibinfo {author} {\bibfnamefont
  {J.~L.}\ \bibnamefont {Luo}}, \bibinfo {author} {\bibfnamefont {N.~L.}\
  \bibnamefont {Wang}}, \ and\ \bibinfo {author} {\bibfnamefont {P.~C}\
  \bibnamefont {Dai}},\ }\bibfield  {title} {\enquote {\bibinfo {title}
  {Structural and magnetic phase transitions in {N}a$_{1-\delta}${F}e{A}s},}\
  }\href {\doibase 10.1103/PhysRevB.80.020504} {\bibfield  {journal} {\bibinfo
  {journal} {Phys. Rev. B}\ }\textbf {\bibinfo {volume} {80}},\ \bibinfo
  {pages} {020504} (\bibinfo {year} {2009})}\BibitemShut {NoStop}%
\bibitem [{\citenamefont {de~la Cruz}\ \emph
  {et~al.}(2008{\natexlab{b}})\citenamefont {de~la Cruz}, \citenamefont
  {Huang}, \citenamefont {Lynn}, \citenamefont {Li}, \citenamefont {II},
  \citenamefont {Zarestky}, \citenamefont {Mook}, \citenamefont {Chen},
  \citenamefont {Luo}, \citenamefont {Wang},\ and\ \citenamefont
  {Dai}}]{delaCruz2008}%
  \BibitemOpen
  \bibfield  {author} {\bibinfo {author} {\bibfnamefont {C.}~\bibnamefont
  {de~la Cruz}}, \bibinfo {author} {\bibfnamefont {Q.}~\bibnamefont {Huang}},
  \bibinfo {author} {\bibfnamefont {J.~W.}\ \bibnamefont {Lynn}}, \bibinfo
  {author} {\bibfnamefont {J.Y.}\ \bibnamefont {Li}}, \bibinfo {author}
  {\bibfnamefont {W.~R.}\ \bibnamefont {II}}, \bibinfo {author} {\bibfnamefont
  {J.~L.}\ \bibnamefont {Zarestky}}, \bibinfo {author} {\bibfnamefont {H.~A.}\
  \bibnamefont {Mook}}, \bibinfo {author} {\bibfnamefont {G.~F.}\ \bibnamefont
  {Chen}}, \bibinfo {author} {\bibfnamefont {J.~L.}\ \bibnamefont {Luo}},
  \bibinfo {author} {\bibfnamefont {N.~L.}\ \bibnamefont {Wang}}, \ and\
  \bibinfo {author} {\bibfnamefont {P.~C}\ \bibnamefont {Dai}},\ }\bibfield
  {title} {\enquote {\bibinfo {title} {Magnetic order close to
  superconductivity in the iron-based layered {L}a{O}$_{1-x}${F}$_x${F}e{A}s
  systems},}\ }\href {\doibase 10.1038/nature07057} {\bibfield  {journal}
  {\bibinfo  {journal} {Nature}\ }\textbf {\bibinfo {volume} {453}},\ \bibinfo
  {pages} {899} (\bibinfo {year} {2008}{\natexlab{b}})}\BibitemShut {NoStop}%
\bibitem [{\citenamefont {McQueen}\ \emph {et~al.}(2009)\citenamefont
  {McQueen}, \citenamefont {Williams}, \citenamefont {Stephens}, \citenamefont
  {Tao}, \citenamefont {Zhu}, \citenamefont {Ksenofontov}, \citenamefont
  {Casper}, \citenamefont {Felser},\ and\ \citenamefont
  {Cava}}]{McQueen2009PRL}%
  \BibitemOpen
  \bibfield  {author} {\bibinfo {author} {\bibfnamefont {T.~M.}\ \bibnamefont
  {McQueen}}, \bibinfo {author} {\bibfnamefont {A.~J.}\ \bibnamefont
  {Williams}}, \bibinfo {author} {\bibfnamefont {P.~W.}\ \bibnamefont
  {Stephens}}, \bibinfo {author} {\bibfnamefont {J.}~\bibnamefont {Tao}},
  \bibinfo {author} {\bibfnamefont {Y.}~\bibnamefont {Zhu}}, \bibinfo {author}
  {\bibfnamefont {V.}~\bibnamefont {Ksenofontov}}, \bibinfo {author}
  {\bibfnamefont {F.}~\bibnamefont {Casper}}, \bibinfo {author} {\bibfnamefont
  {C.}~\bibnamefont {Felser}}, \ and\ \bibinfo {author} {\bibfnamefont {R.~J.}\
  \bibnamefont {Cava}},\ }\bibfield  {title} {\enquote {\bibinfo {title}
  {Tetragonal-to-orthorhombic structural phase transition at 90 {K} in the
  superconductor {F}e$_{1.01}${S}e},}\ }\href {\doibase
  10.1103/PhysRevLett.103.057002} {\bibfield  {journal} {\bibinfo  {journal}
  {Phys. Rev. Lett.}\ }\textbf {\bibinfo {volume} {103}},\ \bibinfo {pages}
  {057002} (\bibinfo {year} {2009})}\BibitemShut {NoStop}%
\bibitem [{\citenamefont {Wang}\ \emph {et~al.}(2010)\citenamefont {Wang},
  \citenamefont {Liu}, \citenamefont {Lv}, \citenamefont {Deng}, \citenamefont
  {Zhao}, \citenamefont {Yu}, \citenamefont {Zhu},\ and\ \citenamefont
  {Jin}}]{WangXC2010LiFeAs}%
  \BibitemOpen
  \bibfield  {author} {\bibinfo {author} {\bibfnamefont {X.~C.}\ \bibnamefont
  {Wang}}, \bibinfo {author} {\bibfnamefont {Q.~Q.}\ \bibnamefont {Liu}},
  \bibinfo {author} {\bibfnamefont {Y.~X.}\ \bibnamefont {Lv}}, \bibinfo
  {author} {\bibfnamefont {Z.}~\bibnamefont {Deng}}, \bibinfo {author}
  {\bibfnamefont {K.}~\bibnamefont {Zhao}}, \bibinfo {author} {\bibfnamefont
  {R.~C.}\ \bibnamefont {Yu}}, \bibinfo {author} {\bibfnamefont {J.~L.}\
  \bibnamefont {Zhu}}, \ and\ \bibinfo {author} {\bibfnamefont {C.~Q}\
  \bibnamefont {Jin}},\ }\bibfield  {title} {\enquote {\bibinfo {title}
  {Superconducting properties of ``111'' type {L}i{F}e{A}s iron arsenide single
  crystals},}\ }\href {\doibase 10.1007/s11433-010-4031-0} {\bibfield
  {journal} {\bibinfo  {journal} {Sci. China Phys. Mech.}\ }\textbf {\bibinfo
  {volume} {53}},\ \bibinfo {pages} {1199} (\bibinfo {year}
  {2010})}\BibitemShut {NoStop}%
\bibitem [{\citenamefont {Cardona}(1983)}]{CardonaBookI}%
  \BibitemOpen
  \bibfield  {author} {\bibinfo {author} {\bibfnamefont {K.}~\bibnamefont
  {Cardona}},\ }\bibfield  {title} {\enquote {\bibinfo {title} {Light
  scattering in solids {I}, {I}ntroductory {C}oncepts},}\ }\href
  {https://link.springer.com/book/10.1007\%2F3-540-11913-2} {\bibfield
  {journal} {\bibinfo  {journal} {Topics in Applied Physics}\ }\textbf
  {\bibinfo {volume} {8}} (\bibinfo {year} {1983})}\BibitemShut {NoStop}%
\bibitem [{\citenamefont {B\"{o}hmer}\ \emph {et~al.}({2013})\citenamefont
  {B\"{o}hmer}, \citenamefont {Hardy}, \citenamefont {Eilers}, \citenamefont
  {Ernst}, \citenamefont {Adelmann}, \citenamefont {Schweiss}, \citenamefont
  {Wolf},\ and\ \citenamefont {Meingast}}]{BohmerPRB87}%
  \BibitemOpen
  \bibfield  {author} {\bibinfo {author} {\bibfnamefont {A.~E.}\ \bibnamefont
  {B\"{o}hmer}}, \bibinfo {author} {\bibfnamefont {F.}~\bibnamefont {Hardy}},
  \bibinfo {author} {\bibfnamefont {F.}~\bibnamefont {Eilers}}, \bibinfo
  {author} {\bibfnamefont {D.}~\bibnamefont {Ernst}}, \bibinfo {author}
  {\bibfnamefont {P.}~\bibnamefont {Adelmann}}, \bibinfo {author}
  {\bibfnamefont {P.}~\bibnamefont {Schweiss}}, \bibinfo {author}
  {\bibfnamefont {T.}~\bibnamefont {Wolf}}, \ and\ \bibinfo {author}
  {\bibfnamefont {C.}~\bibnamefont {Meingast}},\ }\bibfield  {title} {\enquote
  {\bibinfo {title} {{Lack of coupling between superconductivity and
  orthorhombic distortion in stoichiometric single-crystalline FeSe}},}\ }\href
  {\doibase 10.1103/PhysRevB.87.180505} {\bibfield  {journal} {\bibinfo
  {journal} {{Phys. Rev. B}}\ }\textbf {\bibinfo {volume} {{87}}},\ \bibinfo
  {pages} {{180505(R)}} (\bibinfo {year} {{2013}})}\BibitemShut {NoStop}%
\bibitem [{\citenamefont {B\"ohmer}\ \emph {et~al.}({2015})\citenamefont
  {B\"ohmer}, \citenamefont {Arai}, \citenamefont {Hardy}, \citenamefont
  {Hattori}, \citenamefont {Iye}, \citenamefont {Wolf}, \citenamefont
  {v~Loehneysen}, \citenamefont {Ishida},\ and\ \citenamefont
  {Meingast}}]{BohmerPRL114}%
  \BibitemOpen
  \bibfield  {author} {\bibinfo {author} {\bibfnamefont {A.~E.}\ \bibnamefont
  {B\"ohmer}}, \bibinfo {author} {\bibfnamefont {T.}~\bibnamefont {Arai}},
  \bibinfo {author} {\bibfnamefont {F.}~\bibnamefont {Hardy}}, \bibinfo
  {author} {\bibfnamefont {T.}~\bibnamefont {Hattori}}, \bibinfo {author}
  {\bibfnamefont {T.}~\bibnamefont {Iye}}, \bibinfo {author} {\bibfnamefont
  {T.}~\bibnamefont {Wolf}}, \bibinfo {author} {\bibfnamefont {H.}~\bibnamefont
  {v~Loehneysen}}, \bibinfo {author} {\bibfnamefont {K.}~\bibnamefont
  {Ishida}}, \ and\ \bibinfo {author} {\bibfnamefont {C.}~\bibnamefont
  {Meingast}},\ }\bibfield  {title} {\enquote {\bibinfo {title} {{Origin of the
  Tetragonal-to-Orthorhombic Phase Transition in FeSe: A Combined Thermodynamic
  and NMR Study of Nematicity}},}\ }\href {\doibase
  10.1103/PhysRevLett.114.027001} {\bibfield  {journal} {\bibinfo  {journal}
  {{Phys. Rev. Lett.}}\ }\textbf {\bibinfo {volume} {{114}}},\ \bibinfo {pages}
  {{027001}} (\bibinfo {year} {{2015}})}\BibitemShut {NoStop}%
\bibitem [{\citenamefont {Litvinchuk}\ \emph {et~al.}(2008)\citenamefont
  {Litvinchuk}, \citenamefont {Hadjiev}, \citenamefont {Iliev}, \citenamefont
  {Lv}, \citenamefont {Guloy},\ and\ \citenamefont
  {Chu}}]{Litvinchuk2008PhysRevB}%
  \BibitemOpen
  \bibfield  {author} {\bibinfo {author} {\bibfnamefont {A.~P.}\ \bibnamefont
  {Litvinchuk}}, \bibinfo {author} {\bibfnamefont {V.~G.}\ \bibnamefont
  {Hadjiev}}, \bibinfo {author} {\bibfnamefont {M.~N.}\ \bibnamefont {Iliev}},
  \bibinfo {author} {\bibfnamefont {Bing}\ \bibnamefont {Lv}}, \bibinfo
  {author} {\bibfnamefont {A.~M.}\ \bibnamefont {Guloy}}, \ and\ \bibinfo
  {author} {\bibfnamefont {C.~W.}\ \bibnamefont {Chu}},\ }\bibfield  {title}
  {\enquote {\bibinfo {title} {Raman-scattering study of
  {K}$_{x}${S}r$_{1-x}${F}e$_{2}${A}s$_{2}$($x=$0.0,0.4)},}\ }\href {\doibase
  10.1103/PhysRevB.78.060503} {\bibfield  {journal} {\bibinfo  {journal} {Phys.
  Rev. B}\ }\textbf {\bibinfo {volume} {78}},\ \bibinfo {pages} {060503}
  (\bibinfo {year} {2008})}\BibitemShut {NoStop}%
\bibitem [{\citenamefont {Chauvi\`ere}\ \emph {et~al.}(2009)\citenamefont
  {Chauvi\`ere}, \citenamefont {Gallais}, \citenamefont {Cazayous},
  \citenamefont {Sacuto}, \citenamefont {M\'easson}, \citenamefont {Colson},\
  and\ \citenamefont {Forget}}]{Chauviere2009PhysRevB80}%
  \BibitemOpen
  \bibfield  {author} {\bibinfo {author} {\bibfnamefont {L.}~\bibnamefont
  {Chauvi\`ere}}, \bibinfo {author} {\bibfnamefont {Y.}~\bibnamefont
  {Gallais}}, \bibinfo {author} {\bibfnamefont {M.}~\bibnamefont {Cazayous}},
  \bibinfo {author} {\bibfnamefont {A.}~\bibnamefont {Sacuto}}, \bibinfo
  {author} {\bibfnamefont {M.~A.}\ \bibnamefont {M\'easson}}, \bibinfo {author}
  {\bibfnamefont {D.}~\bibnamefont {Colson}}, \ and\ \bibinfo {author}
  {\bibfnamefont {A.}~\bibnamefont {Forget}},\ }\bibfield  {title} {\enquote
  {\bibinfo {title} {Doping dependence of the lattice dynamics in
  {B}a({F}e$_{1-x}${C}o$_{x}$)$_{2}${A}s$_{2}$ studied by {R}aman
  spectroscopy},}\ }\href {\doibase 10.1103/PhysRevB.80.094504} {\bibfield
  {journal} {\bibinfo  {journal} {Phys. Rev. B}\ }\textbf {\bibinfo {volume}
  {80}},\ \bibinfo {pages} {094504} (\bibinfo {year} {2009})}\BibitemShut
  {NoStop}%
\bibitem [{\citenamefont {Rahlenbeck}\ \emph {et~al.}(2009)\citenamefont
  {Rahlenbeck}, \citenamefont {Sun}, \citenamefont {Sun}, \citenamefont {Lin},
  \citenamefont {Keimer},\ and\ \citenamefont {Ulrich}}]{Rahlenbeck2009PRB}%
  \BibitemOpen
  \bibfield  {author} {\bibinfo {author} {\bibfnamefont {M.}~\bibnamefont
  {Rahlenbeck}}, \bibinfo {author} {\bibfnamefont {G.~L.}\ \bibnamefont {Sun}},
  \bibinfo {author} {\bibfnamefont {D.~L.}\ \bibnamefont {Sun}}, \bibinfo
  {author} {\bibfnamefont {C.~T.}\ \bibnamefont {Lin}}, \bibinfo {author}
  {\bibfnamefont {B.}~\bibnamefont {Keimer}}, \ and\ \bibinfo {author}
  {\bibfnamefont {C.}~\bibnamefont {Ulrich}},\ }\bibfield  {title} {\enquote
  {\bibinfo {title} {Phonon anomalies in pure and underdoped
  {R}$_{1-x}${K}$_{x}${F}e$_{2}${A}s$_{2}$ ({R}={B}a, {S}r) investigated by
  raman light scattering},}\ }\href {\doibase 10.1103/PhysRevB.80.064509}
  {\bibfield  {journal} {\bibinfo  {journal} {Phys. Rev. B}\ }\textbf {\bibinfo
  {volume} {80}},\ \bibinfo {pages} {064509} (\bibinfo {year}
  {2009})}\BibitemShut {NoStop}%
\bibitem [{\citenamefont {Hadjiev}\ \emph {et~al.}(2008)\citenamefont
  {Hadjiev}, \citenamefont {Iliev}, \citenamefont {Sasmal}, \citenamefont
  {Sun},\ and\ \citenamefont {Chu}}]{Hadjiev2008PhysRevB}%
  \BibitemOpen
  \bibfield  {author} {\bibinfo {author} {\bibfnamefont {V.~G.}\ \bibnamefont
  {Hadjiev}}, \bibinfo {author} {\bibfnamefont {M.~N.}\ \bibnamefont {Iliev}},
  \bibinfo {author} {\bibfnamefont {K.}~\bibnamefont {Sasmal}}, \bibinfo
  {author} {\bibfnamefont {Y.-Y.}\ \bibnamefont {Sun}}, \ and\ \bibinfo
  {author} {\bibfnamefont {C.~W.}\ \bibnamefont {Chu}},\ }\bibfield  {title}
  {\enquote {\bibinfo {title} {Raman spectroscopy of {R}{F}e{A}s{O}({R}={S}m,
  {L}a)},}\ }\href {\doibase 10.1103/PhysRevB.77.220505} {\bibfield  {journal}
  {\bibinfo  {journal} {Phys. Rev. B}\ }\textbf {\bibinfo {volume} {77}},\
  \bibinfo {pages} {220505} (\bibinfo {year} {2008})}\BibitemShut {NoStop}%
\bibitem [{\citenamefont {Zhao}\ \emph {et~al.}(2009)\citenamefont {Zhao},
  \citenamefont {Hou}, \citenamefont {Wu}, \citenamefont {Xia}, \citenamefont
  {Zhang}, \citenamefont {Chen}, \citenamefont {Luo}, \citenamefont {Wang},
  \citenamefont {Wei}, \citenamefont {Lu},\ and\ \citenamefont
  {Zhang}}]{Zhao2009SCT}%
  \BibitemOpen
  \bibfield  {author} {\bibinfo {author} {\bibfnamefont {S.~C.}\ \bibnamefont
  {Zhao}}, \bibinfo {author} {\bibfnamefont {D.}~\bibnamefont {Hou}}, \bibinfo
  {author} {\bibfnamefont {Y.}~\bibnamefont {Wu}}, \bibinfo {author}
  {\bibfnamefont {T.~L.}\ \bibnamefont {Xia}}, \bibinfo {author} {\bibfnamefont
  {A.~M.}\ \bibnamefont {Zhang}}, \bibinfo {author} {\bibfnamefont {G.~F.}\
  \bibnamefont {Chen}}, \bibinfo {author} {\bibfnamefont {J.~L.}\ \bibnamefont
  {Luo}}, \bibinfo {author} {\bibfnamefont {N.~L.}\ \bibnamefont {Wang}},
  \bibinfo {author} {\bibfnamefont {J.~H.}\ \bibnamefont {Wei}}, \bibinfo
  {author} {\bibfnamefont {Z.~Y.}\ \bibnamefont {Lu}}, \ and\ \bibinfo {author}
  {\bibfnamefont {Q.~M.}\ \bibnamefont {Zhang}},\ }\bibfield  {title} {\enquote
  {\bibinfo {title} {Raman spectra in iron-based quaternary
  {C}e{O}$_{1-x}${F}$_x${F}e{A}s and {L}a{O}$_{1-x}${F}$_x${F}e{A}s},}\ }\href
  {http://stacks.iop.org/0953-2048/22/i=1/a=015017} {\bibfield  {journal}
  {\bibinfo  {journal} {Supercond. Sci. Technol.}\ }\textbf {\bibinfo {volume}
  {22}},\ \bibinfo {pages} {015017} (\bibinfo {year} {2009})}\BibitemShut
  {NoStop}%
\bibitem [{\citenamefont {Fano}(1961)}]{Fano1961PhysRev}%
  \BibitemOpen
  \bibfield  {author} {\bibinfo {author} {\bibfnamefont {U.}~\bibnamefont
  {Fano}},\ }\bibfield  {title} {\enquote {\bibinfo {title} {Effects of
  configuration interaction on intensities and phase shifts},}\ }\href
  {\doibase 10.1103/PhysRev.124.1866} {\bibfield  {journal} {\bibinfo
  {journal} {Phys. Rev.}\ }\textbf {\bibinfo {volume} {124}},\ \bibinfo {pages}
  {1866} (\bibinfo {year} {1961})}\BibitemShut {NoStop}%
\bibitem [{Note1()}]{Note1}%
  \BibitemOpen
  \bibinfo {note} {As a result of the sum rule, what we observe in the $X'X'$
  scattering geometry is the sum of the bare mode and the Fano
  interference.}\BibitemShut {Stop}%
\bibitem [{\citenamefont {Zhang}\ \emph
  {et~al.}(2016{\natexlab{b}})\citenamefont {Zhang}, \citenamefont {Sefat},
  \citenamefont {Ding}, \citenamefont {Richard},\ and\ \citenamefont
  {Blumberg}}]{Zhang2016PRB94}%
  \BibitemOpen
  \bibfield  {author} {\bibinfo {author} {\bibfnamefont {W.-L.}\ \bibnamefont
  {Zhang}}, \bibinfo {author} {\bibfnamefont {Athena~S.}\ \bibnamefont
  {Sefat}}, \bibinfo {author} {\bibfnamefont {H.}~\bibnamefont {Ding}},
  \bibinfo {author} {\bibfnamefont {P.}~\bibnamefont {Richard}}, \ and\
  \bibinfo {author} {\bibfnamefont {G.}~\bibnamefont {Blumberg}},\ }\bibfield
  {title} {\enquote {\bibinfo {title} {Stress-induced nematicity in
  {E}u{F}e$_2${A}s$_2$ studied by {R}aman spectroscopy},}\ }\href {\doibase
  10.1103/PhysRevB.94.014513} {\bibfield  {journal} {\bibinfo  {journal} {Phys.
  Rev. B}\ }\textbf {\bibinfo {volume} {94}},\ \bibinfo {pages} {014513}
  (\bibinfo {year} {2016}{\natexlab{b}})}\BibitemShut {NoStop}%
\bibitem [{\citenamefont {Boeri}\ \emph
  {et~al.}(2010{\natexlab{b}})\citenamefont {Boeri}, \citenamefont {Calandra},
  \citenamefont {Mazin}, \citenamefont {Dolgov},\ and\ \citenamefont
  {Mauri}}]{Boeri2010PRB82}%
  \BibitemOpen
  \bibfield  {author} {\bibinfo {author} {\bibfnamefont {L.}~\bibnamefont
  {Boeri}}, \bibinfo {author} {\bibfnamefont {M.}~\bibnamefont {Calandra}},
  \bibinfo {author} {\bibfnamefont {I.~I.}\ \bibnamefont {Mazin}}, \bibinfo
  {author} {\bibfnamefont {O.~V.}\ \bibnamefont {Dolgov}}, \ and\ \bibinfo
  {author} {\bibfnamefont {F.}~\bibnamefont {Mauri}},\ }\bibfield  {title}
  {\enquote {\bibinfo {title} {Effects of magnetism and doping on the
  electron-phonon coupling in {B}a{F}e$_2${A}s$_2$},}\ }\href
  {https://link.aps.org/doi/10.1103/PhysRevB.82.020506} {\bibfield  {journal}
  {\bibinfo  {journal} {Phys. Rev. B}\ }\textbf {\bibinfo {volume} {82}},\
  \bibinfo {pages} {020506} (\bibinfo {year} {2010}{\natexlab{b}})}\BibitemShut
  {NoStop}%
\bibitem [{\citenamefont {Boeri}\ \emph {et~al.}(2008)\citenamefont {Boeri},
  \citenamefont {Dolgov},\ and\ \citenamefont {Golubov}}]{Boeri2008PRL101}%
  \BibitemOpen
  \bibfield  {author} {\bibinfo {author} {\bibfnamefont {L.}~\bibnamefont
  {Boeri}}, \bibinfo {author} {\bibfnamefont {O.~V.}\ \bibnamefont {Dolgov}}, \
  and\ \bibinfo {author} {\bibfnamefont {A.~A.}\ \bibnamefont {Golubov}},\
  }\bibfield  {title} {\enquote {\bibinfo {title} {Is
  {L}a{F}e{A}s{O}$_{1-x}${F}$_{x}$ an electron-phonon superconductor?}}\ }\href
  {\doibase 10.1103/PhysRevLett.101.026403} {\bibfield  {journal} {\bibinfo
  {journal} {Phys. Rev. Lett.}\ }\textbf {\bibinfo {volume} {101}},\ \bibinfo
  {pages} {026403} (\bibinfo {year} {2008})}\BibitemShut {NoStop}%
\bibitem [{\citenamefont {Coh}\ \emph {et~al.}(2016)\citenamefont {Coh},
  \citenamefont {Cohen},\ and\ \citenamefont {Louie}}]{Coh2016PRB}%
  \BibitemOpen
  \bibfield  {author} {\bibinfo {author} {\bibfnamefont {S.}~\bibnamefont
  {Coh}}, \bibinfo {author} {\bibfnamefont {M.~L.}\ \bibnamefont {Cohen}}, \
  and\ \bibinfo {author} {\bibfnamefont {S.~G.}\ \bibnamefont {Louie}},\
  }\bibfield  {title} {\enquote {\bibinfo {title} {Antiferromagnetism enables
  electron-phonon coupling in iron-based superconductors},}\ }\href {\doibase
  10.1103/PhysRevB.94.104505} {\bibfield  {journal} {\bibinfo  {journal} {Phys.
  Rev. B}\ }\textbf {\bibinfo {volume} {94}},\ \bibinfo {pages} {104505}
  (\bibinfo {year} {2016})}\BibitemShut {NoStop}%
\bibitem [{\citenamefont {Coh}\ \emph {et~al.}(2015)\citenamefont {Coh},
  \citenamefont {Cohen},\ and\ \citenamefont {Louie}}]{Coh2015NJP}%
  \BibitemOpen
  \bibfield  {author} {\bibinfo {author} {\bibfnamefont {S.}~\bibnamefont
  {Coh}}, \bibinfo {author} {\bibfnamefont {M.~L.}\ \bibnamefont {Cohen}}, \
  and\ \bibinfo {author} {\bibfnamefont {S.~G.}\ \bibnamefont {Louie}},\
  }\bibfield  {title} {\enquote {\bibinfo {title} {Large electron–phonon
  interactions from {F}e{S}e phonons in a monolayer},}\ }\href
  {http://stacks.iop.org/1367-2630/17/i=7/a=073027} {\bibfield  {journal}
  {\bibinfo  {journal} {New J. Phys.}\ }\textbf {\bibinfo {volume} {17}},\
  \bibinfo {pages} {073027} (\bibinfo {year} {2015})}\BibitemShut {NoStop}%
\end{thebibliography}
\end{document}